\documentclass[10pt,twocolumn,aps,prb,showpacs,superscriptaddress,groupedaddress,citeautoscript]{revtex4-1}
\usepackage{graphicx}  
\usepackage{dcolumn}   
\usepackage{bm}        
\usepackage{amssymb}  
\usepackage{amsmath}

\newcommand{\beq}{\begin{equation}}
\newcommand{\eeq}{\end{equation}}
\newcommand{\bea}[1]{\begin{subequations}\label{#1}\begin{eqnarray}}
\newcommand{\eea}{\end{eqnarray}\end{subequations}}
\newcommand{\bal}{\begin{align}}
\newcommand{\eal}{\end{align}}

\newcommand{\eqlabel}[2]{\begin{equation}\label{#1} #2 \end{equation}}
\newcommand{\eq}[1]{\begin{equation} #1 \end{equation}}
\newcommand{\Mat}[4]{\left(\begin{array}{cc} #1 & #2\\ #3 & #4 \end{array}\right)}

\begin{document}
\title{Effective Medium Theory for Elastic Metamaterials in Thin Elastic Plates}
\author{Daniel Torrent}
\email{daniel.torrent@iemn.univ-lille1.fr}
\author{Yan Pennec}
\author{Bahram Djafari-Rouhani}
\affiliation{Institut d'Electronique, de Microl\'ectronique et de Nanotechnologie, UMR CNRS 8520, Universit\'e de Lille 
1, 59655 Villeneuve d’Ascq, France}

\date{\today}

\begin{abstract}
An effective medium theory for resonant and non-resonant metamaterials for flexural waves in thin plates is presented. The theory provides closed-form expressions for the effective mass density, rigidity and Poisson's ratio of arrangements of isotropic scatterers in thin plates, valid for low frequencies and moderate filling fractions. It is found that the effective Young's modulus and Poisson's ratio are induced by a combination of the monopolar and quadrupolar scattering coefficient, as it happens for bulk elastic waves, while the effective mass density is induced by the monopolar coefficient, contrarily as it happens for bulk elastic waves, where the effective mass density is induced by the dipolar coefficient. It is shown that resonant positive or negative effective elastic parameters are possible, being therefore the monopolar resonance the responsible of creating an effective medium with negative mass density. Several examples are given for both non-resonant and resonant effective parameters and the results are numerically verified by multiple scattering theory and finite element analysis.
\end{abstract}
\maketitle
\section{Introduction}
In the last few years the field of metamaterials has received increasing attention, due to the extraordinary properties of these structures to control the propagation characteristics of electromagnetic\cite{smith2004metamaterials,pendry2006controlling}, acoustic\cite{li2004double,cummer2007one} or elastic \cite{liu2000locally,milton2006cloaking} waves. Consisting essentially in periodic arrangements of wave interacting
units or scatterers, these advanced structures behave as materials with extraordinary constitutive parameters, like negative permittivity\cite{schurig2006electric}, anisotropic mass density\cite{torrent2008anisotropic} or negative elastic modulus\cite{fang2006ultrasonic}. The wide variety of phenomena and applications found for these structures has motivated the research in this field not only for bulk waves but also for confined\cite{torrent2009radial,torrent2010acoustic} or surface waves\cite{christensen2007collimation,torrent2012acoustic}.

More specifically, the control of the propagation of flexural waves in thin elastic plates has been widely studied. For instance, cloaking devices \cite{Farhat2009,Farhat2009a,Stenger2012}, flat lenses based on negative refraction\cite{Farhat2010,pierre2010negative}, gradient index flat and circular lenses\cite{lenteTTWu,zhao2012efficient,climente2014gradient} and omnidirectional absorbers\cite{krylov2012resumen,climente2013omnidirectional} for flexural waves have been recently proposed and experimentally verified. Also, the dispersion relation of plates with periodic arrangements of rigid pins\cite{Evans2007,mcphedran2009platonic}, holes\cite{Movchan2007}, attached pillars \cite{pennec2008low,pennec2009phonon,marchal2012dynamics} or point-like spring-mass resonators\cite{Xiao2011,Xiao2012,Torrent2013Graphene} attached to them has been investigated by several groups, as well as the resonant properties of complex inclusions\cite{Zhu2012}.

In the aforementioned works, the properties of the complex structure are mainly determined by band structure calculation, since an effective medium theory for the description of these new metamaterials, specially relating the different resonant parameters with the corresponding symmetry of the field, has not been given so far, as has been done for acoustic or bulk elastic waves. 

In this work we present a theory for elastic metamaterials for flexural waves based on the scattering properties of arrangements of scatterers or resonators. This homogenization method has been previously employed for either electromagnetic\cite{vynck2009all,torrent2011multiple}, acoustic\cite{li2004double} and elastic waves\cite{wu2007effective,zhou2009analytic}, and the resonant effective parameters have been properly assigned to different symmetries of the fields; in this work we obtain similar relationships for flexural waves, but with important differences given that these waves are not described by the Helmholtz equation but by the biharmonic equation\cite{Graff, Timoshenko}.

Thus, the presented theory obtains the effective parameters as a function of the filling fraction of the inclusions, their physical properties and the frequency. The theory is valid for low and mid filling fractions, being necessary the inclusion of the multiple scattering terms in order to cover the full range of inclusions' radii, however it still provides a good description of the effective materials' properties for both frequency dependent and non-dependent structures.

The paper is organized as follows: After this introduction, in Section \ref{sec:homo} the general procedure for obtaining the effective parameters from
the scattering properties of clusters of inclusions or resonators is explained. Following, Section \ref{sec:scattering} describes the scattering properties
of a circular inclusion in a thin plate, and the low frequency behaviour of these scattering properties are analysed  by means of the so called $T$ matrix. This
low-frequency behaviour is used in Section \ref{sec:effective} to obtain closed form expressions for the effective parameters of arrangements of scatterers and resonators in thin plates. The theory is numerically validated in Section \ref{sec:numval}, where two dimensional multiple scattering simulations and three dimensional finite element calculations are performed. Finally, results are summarized in Section \ref{sec:summary}. Appendix \ref{sec:XY} contains some mathematical details about the analytical derivations.

\section{Homogenization from the Scattering Properties}
\label{sec:homo}
This section shows the general procedure to obtain the effective properties of a given collection of scatterers when interacting with a field which satisfies Helmholtz equation. As mentioned in the previous section, this method has been widely employed for the homogenization at low filling fractions of electromagnetic, elastic and acoustic metamaterials. The objective of this section is to present it in such a way that its application to flexural waves be straightforward.

\begin{figure}
\centering
\includegraphics[scale=0.8]{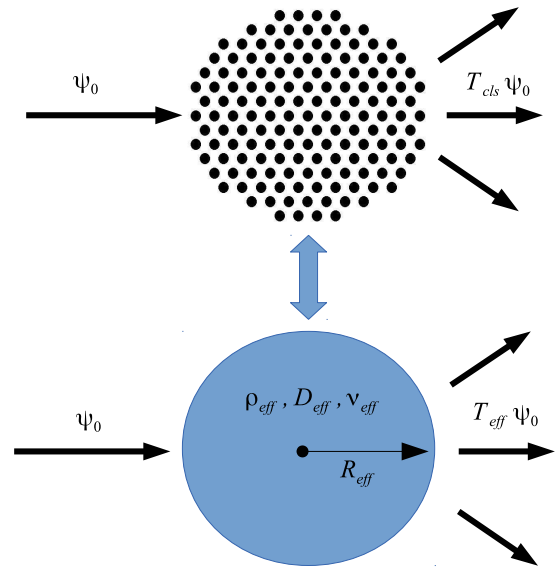}
\caption{\label{fig:schematics} Schematic view of the homogenization process based on the effective $T$ matrix formalism. An incident field $\psi_0$ arrives to a circular cluster
of $N$ scatterers. It is expected that in the low frequency the cluster behaves like an effective circular scatterer of radius $R_{eff}$ and parameters $\rho_{eff},D_{eff}$ and $\nu_{eff}$. The parameters are obtained by comparing the low frequency limit of the scattering properties, defined by means of the cluster and effective $T$ matrices.}
\end{figure}

Figure \ref{fig:schematics} shows the schematic view of a general homogenization procedure based on multiple scattering: An incident field $\psi_0$ impinges a 
circular cluster of scatterers, and it excites a scattered field $\psi_{sc}=T_{cls}\psi_0$, being $T_{cls}$ the $T$ matrix of the cluster, which is a function of the frequency and the physical properties of the scatterers. We expect that, in the low frequency limit, this cluster behaves
like an effective circular scatterer of radius $R_{eff}$ and some effective parameters which will depend on the nature of the wave under study. In this work,
as will be shown later, the parameters describing any scatterer will be the effective mass density $\rho_{eff}$, the effective Poisson's ratio $\nu_{eff}$ and 
the effective rigidity $D_{eff}$. 

The homogenization procedure consists in obtaining these effective parameters from the low frequency behaviour of the $T$ matrix of the cluster, since it is expected to be identical to the $T$ matrix of the effective scatterer $T_{eff}$ computed by means of the effective parameters to be obtained. Thus, the effective parameters are obtained from \cite{homoDani2}
\eq{
\lim_{\omega\to 0}T_{cls}=\lim_{\omega\to 0}T_{eff}.
}
The left hand side of the above equation is a known quantity, since we decide the nature, size and position of the scatterers, while the right hand side  of the equation contains the parameters to be determined, therefore the above equation provides a solution for the effective parameters of the medium.

In general, in either two and three dimensions, both the incident field and the scattered field are expanded in multipolar fields, for instance, in two dimensions we have that the fields are in general given by
\bea{eq:psi0SC}
\psi_0&=&\sum_q A_q \psi_q^0(k_br)e^{iq\theta}\\
\psi_{sc}&=&\sum_q B_q \psi_q^{sc}(k_br)e^{iq\theta},
\eea 
being $r$ and $\theta$ the polar coordinates and $q$ an integer number representing the mode's symmetry. Here, $k_b$ is the background's wave number and $\psi_q^0$ and $\psi_q^{sc}$ are usually Bessel and Hankel (or related) functions representing the incident and scattered fields, respectively. The $T$ matrix  relates the $B_q$ coefficients with the $A_q$, that is
\eqlabel{eq:BqTqsAs}{
B_q=\sum_s T_{qs}A_s.
}

In the case of a circular homogeneous scatterer, the $T$ matrix becomes diagonal, then $T_{qs}=T_q\delta_{qs}$. The $T$ matrix of a cluster of scatterers is not diagonal in general, since it is obvious that the structure is not invariant under rotations, but in the low frequency limit this $T$ matrix becomes diagonal\cite{homoDani2}, since in this limit it is behaving like a homogeneous scatterer (actually this is only true for isotropic arrangements of scatterers, the anisotropic case is different and beyond the scope of the present work). 

In the low frequency limit all the $T_q$ elements goes to zero, although the trend is different for each multipolar order $q$. For instance, in the case of electromagnetic or acoustic scatterers, the dominant terms are the $T_0$ and $T_1$ elements, which goes to zero as $\omega^2$, while for elastic waves in solids the dominant
terms are the $T_0,T_1$ and $T_2$ elements. It is from the dominant terms from which we obtain the effective parameters, and the number of dominant terms is
consistent with the number of parameters to determine. Thus, for electromagnetic waves we need two dominant terms, since we have to obtain the effective 
permeability $\mu$  and permittivity $\varepsilon$. Similarly, in acoustics we need to obtain the effective bulk modulus $B$ and mass density $\rho$ and again
we have two dominant terms. Finally, for elastic waves in solids, we need three parameters, the two Lam\'e coefficients $\lambda$ and $\mu$ and the mass density, 
and we have three dominant terms in the low frequency limit.

It can be shown\cite{waterman1965matrix,waterman2005new} that the general structure of the dominant term of the $T$ matrix of a homogeneous scatterer of radius $R_a$ is, in two dimensions,
\eq{
T_q\approx \frac{i\pi k_b^2R_a^2}{4}\Gamma_q^{a},
}
being $\Gamma_q^a$ a function of the scatterer's parameters, whose explicit expression depends on the nature of the field considered. The expression is valid as well for the effective scatterer, therefore in this limit the dominant terms of the effective $T$ matrix are
\eq{
(T_q)_{eff}\approx \frac{i\pi k_b^2R_{eff}^2}{4}\Gamma_q^{eff}.
}
Also, it can be shown that the $T$ matrix of the cluster is, as a first approximation, i.e., neglecting the multiple scattering terms, the addition of the $T$ matrices of all the scatterers\cite{homoDani2}. If all them are identical and we have $N$ scatterers in the cluster we get
\eq{
(T_q)_{cls}\approx N\frac{i\pi k_b^2R_a^2}{4}\Gamma_q^{a}.
}

The homogenization procedure implies equating the last two equations and defining the filling fraction as 
\beq
f=NR_a^2/R_{eff}^2,
\eeq
then we get
\eqlabel{eq:chieff}{
\Gamma_q^{eff}=f\Gamma_q^a,
}
from which we can obtain the effective parameters. If instead of having $N$ identical scatterers we have $N_1$ scatterers of type 1, $N_2$ scatterers of type 2 and
so on, and they form an isotropic medium, we have that
\eqlabel{eq:chieffi}{
\Gamma_q^{eff}=\sum_i f_i\Gamma_q^i,
}
where $f_i$ is the partial filling fraction of the scatterer of type $i$.

It must be pointed out that the filling fraction is defined here in terms of the effective radius of the cluster $R_{eff}$, which, as shown in reference \onlinecite{felbacq1994scattering}, is not well defined and its definition can obviously affect the effective parameters. However, in this work we define the effective radius of the cluster so that the filling fraction be identical to that of the corresponding underlying lattice in which the scatterers are arranged. This definition was employed and analysed by Torrent {\it et al.} in references \onlinecite{homoDani1,homoDani2,torrent2007evidence}, where it was shown that it is a very consistent definition for the analysis of both ordered and disordered systems. For instance, for scatterers arranged in a triangular lattice, the filling fraction is $f=2\pi/\sqrt{3} R_a^2/a^2$, so that the effective radius of the cluster will be given by
\beq
R_{eff}=\sqrt{\frac{\sqrt{3}N}{2\pi}}a.
\eeq

Equations \eqref{eq:chieff} and \eqref{eq:chieffi} does not contain information about the specific order of the scatterers in the cluster. However, since we have assumed that the effective medium is isotropic, we are implicitly limiting the theory to symmetric lattices (triangular or square), being the difference
between these arrangements the definition of the filling fraction. A deeper influence of the lattice symmetry as well as the possibility of having non-symmetric
lattices can be done by the inclusion of the multiple scattering terms and by assuming that the effective scatterer is anisotropic, however this analysis is beyond the objective of the present work (see references \onlinecite{homoDani1,torrent2008anisotropic} for an example of these methods in acoustics).

As an application of the previous method for acoustic waves, we have for instance that $\Gamma_0^a=1-B_b/B_a$, being $B_b$ and $B_a$ the bulk modulus of 
the background and the scatterer, respectively. Using the above equation we obtain the well know expression for the effective bulk modulus of a composite
\eq{
\frac{1}{B_{eff}}=\frac{1-f}{B_b}+\frac{f}{B_a},
}
or, in the most general case,
\eq{
\frac{1}{B_{eff}}=\frac{1-f}{B_b}+\sum_i\frac{f_i}{B_i}.
}

In the next section the scattering of flexural waves in thin plates is described in terms of the $T$ matrix, and the dominant terms are derived for later use in the
extraction of the effective parameters of an ensemble of scatterers.
\section{Scattering of Flexural Waves by a Circular Inclusion}
\label{sec:scattering}
The equations describing flexural waves can be found in many textbooks (see for instance references \onlinecite{Timoshenko,Graff}). If the wavelength of the field is larger than the thickness of the plate, the wave equation is the fourth order differential equation 
\begin{multline}
\label{eq:wave}
-\frac{\partial^2}{\partial x^2}\left(D_b\left[\frac{\partial^2 W}{\partial x^2}+\nu_b\frac{\partial^2 W}{\partial y^2}\right]\right)\\
-\frac{\partial^2}{\partial y^2}\left(D_b\left[\frac{\partial^2 W}{\partial y^2}+\nu_b\frac{\partial^2 W}{\partial x^2}\right]\right)\\
-2\frac{\partial^2}{\partial x\partial y}\left(D_b(1-\nu_b)\frac{\partial^2 W}{\partial x\partial y}\right)=\rho h\frac{\partial^2 W}{\partial t^2},
\end{multline}
being $\rho_b$, $h_b$ and $D_b=E_bh_b^3/12(1-\nu_b^2)$ the mass density, thickness and rigidity of the plate, respectively, with $E_b$ the Young's modulus and $\nu_b$  and Poisson's ratio. When the background's parameters are constant and we assume harmonic time dependence of the field $W$, the above equation reduces to
\eq{
(D_b\nabla^4-\rho_bh_b\omega^2)W(x,y)=0,
}
whose solution in polar coordinates is given by a linear combination of Bessel and modified Bessel functions\cite{NorrisVemula} of argument $k_b$, such
that
\eqlabel{eq:kb}{
k_b^4=\frac{\rho_bh_b}{D_b}\omega^2.
}

In this work, we are interested in describing scattering processes by circular inclusions in the low frequency limit, defined for wavelengths such that $\lambda>4a$, being $a$ the typical distance between scatterers. Thus, as long as the thickness of the plate be smaller than $a$, the above equation is a good approximation. We have chosen the thickness of the plate $h_b=0.1a$, which ensures that the previous theory is valid even when the field propagates inside complex scatterers, as will be discussed in Section \ref{sec:effective}.

For a scattering problem, the incident field is expressed as
\eq{
W_0=\sum_q\left[A_q^JJ_q(k_br)+A_q^II_q(k_br)\right]e^{iq\theta},
}
while the scattered field is given by
\eq{
W_{sc}=\sum_q\left[B_q^HH_q(k_br)+B_q^KK_q(k_br)\right]e^{iq\theta}.
}

If the scatterer is a circular inhomogeneity of radius $R_a$ we have that, inside the scatterer ($r<R_a$), since there are no sources, the field is expressed as
\eq{
W_i=\sum_q\left[C_q^JJ_q(k_ar)+C_q^II_q(k_ar)\right]e^{iq\theta}
}

Boundary conditions are explained for instance in reference \onlinecite{NorrisVemula}, and they provide a system of four equations which solves for the fourth unknowns: two scattering coefficients $B_q^H,B_q^K$ and the two internal coefficients $C_q^J, C_q^I$. The system of equations can be expressed as (see Appendix \ref{sec:XY} for details)
\bea{eq:ABC}
\bm{X}^0_qA_q+\bm{X}^{sc}_qB_q&=\bm{X}^a_qC_q\label{eq:BC1},\\
\bm{Y}^0_qA_q+\bm{Y}^{sc}_qB_q&=\bm{Y}^a_qC_q\label{eq:BC2},
\eea
where the matrices $\bm{X}_q^i$ and $\bm{Y}_q^i$, with $i=0,sc,a$, are $2\times 2$ matrices given in Appendix \ref{sec:XY}
and the coefficient vectors are $A_q=(A_q^J,A_q^I)$, $B_q=(B_q^H,B_q^K)$ and $C_q=(C_q^J,C_q^I)$. Solving for $C_q^i$ from equation \eqref{eq:BC1} and inserting into equation \eqref{eq:BC2} gives
\eq{
\bm{Y}^0_qA_q+\bm{Y}^{sc}_qB_q=\bm{Y}^a_q(\bm{X}^a_q)^{-1}(\bm{X}^0_qA_q+\bm{X}^{sc}_qB_q),
}
from which we can solve for the $B_q$ as a function of $A_q$
\begin{multline}
\label{eq:BqAq}
B_q=-\left(\bm{Y}^{sc}_q-\bm{Y}^a_q(\bm{X}^a_q)^{-1}\bm{X}^{sc}_q\right)^{-1}\times\\
\left(\bm{Y}^0_q-\bm{Y}^a_q(\bm{X}^a_q)^{-1}\bm{X}^0_q\right)A_q.
\end{multline}

The above equation defines the $T$ matrix of the scatterer, and it gives the scattering coefficients $B_q$
as a function of $A_q$. It is a $2x2$ matrix and each element of the matrix relates the excitation of
a different mode, that is, in full matrix form we have
\eq{
\left(\begin{array}{c}
B_q^H \\ B_q^K
\end{array}\right)=
\left(\begin{array}{cc}
T_q^{HJ} & T_q^{HI}\\ T_q^{KJ} & T_q^{KI}
\end{array}\right)
\left(\begin{array}{c}
A_q^J \\ A_q^I
\end{array}\right).
}

If the scatterer is a hole, the clamped free boundary conditions gives simply
\eq{
B_q=-(\bm{Y}_q^{sc})^{-1}\bm{Y}_q^0A_q.
}

The effective medium defined by a cluster of scatterers will be described by means of the three parameters appearing in the wave equation \eqref{eq:wave}, which are the mass density $\rho_{eff}$ (actually the surface mass density $\rho_{eff}h_{eff}$), the effective rigidity $D_{eff}$ and the effective Poisson's ratio $\nu_{eff}$. Thus, in principle we would expect that the dominant terms of the $T$ matrix be three, as it happens for bulk elasticity, however it will be shown that the case of flexural waves is different.

As shown in reference \onlinecite{Parnell2011}, in the low frequency (wavenumber) limit, the elements of the $T$ matrix of a hole depend on each multipolar order $q$, as expected. Thus, the $q=0$ element is
\eq{
T_0\approx \frac{i\pi (k_bR_a)^2}{4}\frac{1}{1-\nu_b}
\begin{pmatrix}
\nu_b & -1 \cr
-2i/\pi  & 2i\nu_b/\pi
\end{pmatrix},
}
while the $q=1$ is
\eq{
T_1\approx \frac{i\pi (k_bR_a)^4}{32}\frac{1}{1-\nu_b}
\begin{pmatrix}
1+\nu_b & -2 \cr
-4i/\pi  & 2i(1+\nu_b)/\pi
\end{pmatrix},
}
finally, for $q\geq2$ we have
\eq{
T_q\approx \frac{i\pi(k_bR_a)^{2q-2}}{2^{2q-1}(q-1)!(q-2)!}\frac{1-\nu_b}{3+\nu_b}
\begin{pmatrix}
1 & 1 \cr
2i/\pi  & 2i/\pi
\end{pmatrix}.
}

The above expressions show that in the low frequency limit the dominant terms of the $T$ matrix are the $q=0$ and the $q=2$, unlike in other waves like acoustic or electromagnetic where the dominant terms are the $q=0$ and the $q=1$. Only in the elastic case we find the $q=2$ as a dominant term, but it also includes the $q=1$. However, although in the case of flexural waves we have found only two dominant terms, and we still have three effective parameters to obtain, the $T$ matrix elements are actually $2\times 2$ matrices, which moreover have the following form
\eqlabel{eq:T0chi0}{
T_0\approx\frac{i\pi (k_b R_a)^2}{4}
\begin{pmatrix}
\Gamma_0^{11} & \Gamma_0^{12} \cr
2i/\pi\Gamma_0^{12} & 2i/\pi \Gamma_0^{11}
\end{pmatrix}
}
and 
\eqlabel{eq:T2chi2}{
T_2\approx\frac{i\pi (k_b R_a)^2}{4}
\begin{pmatrix}
\Gamma_2 & \Gamma_2 \cr
2i/\pi\Gamma_2 & 2i/\pi \Gamma_2
\end{pmatrix}
}
with
\bea{eq:chii}
\Gamma_0^{11}&=&\frac{\nu_b}{1-\nu_b}\\
\Gamma_0^{12}&=&-\frac{1}{1-\nu_b}\\
\Gamma_2&=&\frac{1}{2}\frac{1-\nu_b}{3+\nu_b},
\eea
which shows that indeed we have only three independent terms. This structure of the $T$ matrix should be maintained for a general inhomogeneity, since it
is from that expression from which we will obtain the effective parameters. The demonstration is tedious and long, and some details are given in Appendix \ref{sec:XY}, but as expected the dominant terms of the $T$ matrix of an elastic inhomogeneity have the same behaviour and form, and it is found that
\begin{align}
\Gamma_0^{11}&=\frac{1}{2}\frac{\rho_a h_a}{\rho_b h_b}+\frac{D_b}{D_b(1-\nu_b)+D_a(1+\nu_a)}-1\\
\Gamma_0^{12}&=\frac{1}{2}\frac{\rho_a h_a}{\rho_b h_b}-\frac{D_b}{D_b(1-\nu_b)+D_a(1+\nu_a)}\\
\Gamma_2&=\frac{1}{2}\frac{D_b(1-\nu_b)-D_a(1-\nu_a)}{D_b(3+\nu_b)+D_a(1-\nu_a)}.
\end{align}

Notice that we recover the expressions for the holes by setting $D_a=0$ and $\rho_a=0$. From the above expressions it is now possible to obtain the effective
parameters for an ensemble of scatterers in a thin plate. These parameters are derived in next section, first for the low frequency limit and, later on, the 
generalization for a resonant medium.
\section{Effective Parameters}
\label{sec:effective}
The low frequency limit of the $T$ matrix of a inhomogeneity in a thin plate responds to the general behaviour explained before, therefore application 
of equation \eqref{eq:chieff} is straightforward, giving
\bea{eq:chieffchia}
(\Gamma_0^{11})_{eff}&=&f\Gamma_0^{11}\\
(\Gamma_0^{12})_{eff}&=&f\Gamma_0^{12}\\
(\Gamma_2)_{eff}&=&f \Gamma_2.
\eea
From the above equations we can solve for the effective parameters as a function of the filling fraction, giving
\begin{subequations}
\label{eq:effparams}
\begin{align}
\rho_{eff}&=(1+f(\Gamma_0^{11}+\Gamma_0^{12}))\rho_b\\
D_{eff}(1+\nu_{eff})&=\frac{1+\nu_b-f(\Gamma_0^{11}-\Gamma_0^{12})(1-\nu_b)}{1+f(\Gamma_0^{11}-\Gamma_0^{12})}D_b\\
D_{eff}(1-\nu_{eff})&=\frac{1-\nu_b-2f\Gamma_2(3+\nu_b)}{1+2f\Gamma_2}D_b.
\end{align}
\end{subequations}

The above equations show that the effective mass density is obtained from the monopolar scattering term. When this term be resonant, as will be shown later, we will have an effective medium with a dispersive effective mass density, reaching positive and negative values. This result is remarkably different from that found for bulk elastic waves or acoustic waves, where the effective mass density is induced by the dipolar scattering term. However, it must be pointed out that for flexural waves the mass density relates the shear force with the vertical displacement of the plate, so that the physical origin of this scattering coefficient is clearly different than for bulk elastic waves.

It is clear as well that the effective rigidity and Poisson's ratio are defined from the monopolar and quadrupolar scattering elements, which also happens for bulk elasticity.

\subsection{Non-Resonant Effective Parameters}
The effective parameters for a non-resonant medium are obtained by inserting the low frequency parameters $\Gamma_i$ into equations \eqref{eq:effparams}, from which we obtain the following
expressions
\begin{widetext}
\begin{subequations}
\label{eq:nonresonant}
\begin{align}
\rho_{eff}&=(1-f)\rho_b+f\rho_a\\
D_{eff}(1+\nu_{eff})&=\frac{(1+\nu_b)(D_b(1-\nu_b)+D_a(1+\nu_a))-f(1-\nu_b)(D_b(1+\nu_b)-D_a(1+\nu_a))}{D_b(1-\nu_b)+D_a(1+\nu_a)-f(D_b(1+\nu_b)-D_a(1+\nu_a)}D_b\\
D_{eff}(1-\nu_{eff})&=\frac{(1-\nu_b)(D_b(3+\nu_b)+D_a(1-\nu_a))-f(3+\nu_b)(D_b(1-\nu_b)-D_a(1-\nu_a))}{D_b(3+\nu_b)+D_a(1-\nu_a)-f(D_b(1-\nu_b)-D_a(1-\nu_a))}D_b.
\end{align}
\end{subequations}
\end{widetext}

Notice that the equation for the effective mass density is identical than that found for bulk elastic waves, however here it has been obtained from the monopolar term while in elasticity it is obtained from the dipolar one.

The dispersion relation for a homogeneous medium given by \eqref{eq:kb} shows that the phase velocity is frequency-dependent for flexural waves, however, from the point of view of refraction in stationary problems, it is more interesting the relative phase velocity between the effective medium and the background, which is given by 
\eq{
c_{eff}/c_b=k_b/k_{eff}=(D_{eff}/\rho_{eff})^{1/4}/(D_{b}/\rho_{b})^{1/4},
}
where it has been used that $k_{eff}^4=\rho_{eff}h_{eff}/D_{eff}\omega^2$ and the fact that the effective thickness of the medium does not change, that is, $h_{eff}=h_b$.

Special mention deserves the system of holes in an elastic plate. As said before, the expressions for the
holes are obtained by setting $\rho_a=0$ and $D_a=0$. It is easy to see that then the the effective phase velocity $c_{eff}$ related with the phase velocity of the background $c_b$ depends only on the plate's Poisson's ratio. Effectively, in this particular case, equations \eqref{eq:nonresonant} become
\begin{subequations}
\label{eq:effholes}
\begin{align}
\rho_{eff}&=(1-f)\rho_b,\\
D_{eff}(1+\nu_{eff})&=\frac{(1-\nu_b^2)(1-f)}{(1-\nu_b)-f((1+\nu_b)}D_b,\\
D_{eff}(1-\nu_{eff})&=\frac{(1-\nu_b)(3+\nu_b)(1-f)}{(3+\nu_b)-f(1-\nu_b)}D_b,
\end{align}
\end{subequations}
from which it is clear that the ratio $D_{eff}/\rho_{eff}$ depends only on the filling fraction and the background's Poisson's ratio.

\begin{figure}
\centering
\includegraphics[width=8cm]{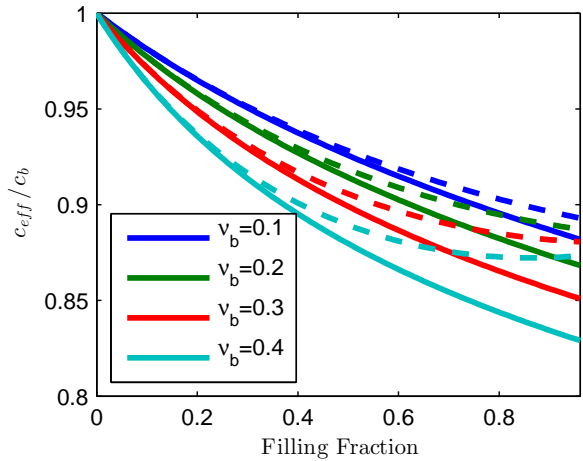}
\caption{\label{fig:ceffholesapp}
Continuous lines:Effective phase velocity for circular holes embedded in an elastic plate of different Poisson's ratio. Dashed lines: Results from an approximated model (see text for further details). }
\end{figure}

Figure \ref{fig:ceffholesapp} shows the effective phase velocity as a function of the filling fraction computed using equations \eqref{eq:nonresonant} for different Poisson's ratio (continuous lines) compared with the second order approximation for the phase velocity given by equation (65) of reference \onlinecite{Parnell2011} (dashed lines). Notice that the the two expressions give the same velocity for low filling fractions but they split for filling fractions above 0.2, where the expression given in the present work is more accurate. It must be also remarked than equations \eqref{eq:nonresonant} give also an approximated expression for high filling fractions, where the multiple scattering terms should be included in the theory\cite{homoDani1,homoDani2}. However, as will be shown in the analysis performed in Section \ref{sec:numval}, these expressions can be properly used for low and mid filling fractions, that is, for $f\lesssim 0.4$.

\begin{figure}
\centering
\includegraphics[width=8cm]{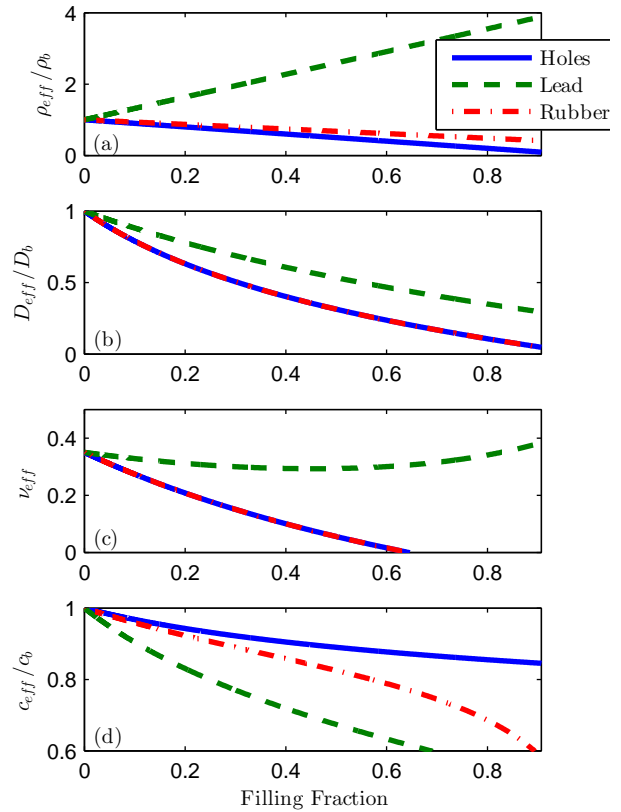}
\caption{\label{fig:effparams}
Effective mass density (a), rigidity (b), Poisson's ratio (c) and phase velocity (d) as a function of the filling fraction for circular inclusions in aluminium. Results are shown for holes (blue continuous line), lead inclusions (green dashed line) and rubber inclusions (red dash-dotted line).}
\end{figure}

Figure \ref{fig:effparams} shows the effective mass density $\rho_{eff}$(panel (a), rigidity $D_{eff}$ (panel (b)), Poisson's ratio $\nu_{eff}$ (panel (c)) and phase velocity $c_{eff}$ (panel (d)) as a function of the filling fraction for circular inclusions of holes, rubber and lead in an aluminium plate, with the elastic parameters given in Table \ref{tab:params}. The effective mass density follows the well known linear relationship with the filling fraction, since it is simply the volume average, as for bulk elastic waves. The effective rigidity $D_{eff}$ has a decreasing trend, being indistinguishable the case of holes to that of rubber due to the low ratio between the Young's modulus of the rubber and the aluminium. The same phenomenon is observed for the Poisson's ratio but not for the phase velocity, where the differences between the rubber and the holes are more evident. 
\begin{table}
\caption{\label{tab:params}Elastic constants of the materials used for the simulations}
\begin{tabular}{|c|c|c|c|}
\hline
 Parameter/Material    & Aluminium & Lead & Rubber \\ \hline \hline
 $\rho\, (Kg/dm^3)$ & 2.71          &   11.34   & 0.96 \\ \hline
	$E\, (GPa)$ & 70      &   16   &7E-4 \\ \hline
  $\nu$ & 0.35          &  0.44    &0.45 \\ \hline
	$D/D_b$ &   1        &  0.64    & 2.9E-5\\ \hline
\end{tabular}
\end{table}

It must be pointed out that the values of the effective Poisson's ratio for the case of holes and the rubber inclusions reach
zero at filling fractions close to 0.6. This is due to the fact that the scatterers are ``very strong'', and 
the multiple scattering corrections must be added for high filling fractions, as explained above. The case
of lead inclusions is well-behaved in this sense, however it does not means that these multiple scattering terms
are not necessary for a proper description. A full analysis of these multiple scattering terms is beyond the scope of the present work, since they are important for strong scatterers and high filling fractions only, however their requirement will be analysed in Section \ref{sec:numval}.



\subsection{Resonant Effective Parameters}
The effective parameters given by equations \eqref{eq:nonresonant} are obtained in the low frequency limit, which implies that the wavenumber in the background, inside the scatterer and in the effective medium are negligible. However, as long as the wavenumber in the background be small (that is, the wavelength larger than the typical distance between scatterers), the description of the system as an effective medium makes sense. It means that we can allow the wavelength inside the scatterer be short and then have complex field oscillations while the field in the background sees an average medium. If this is the case, we will have a frequency-dependent effective medium. Under these conditions the asymptotic expressions for low arguments of Bessel functions cannot be employed for the quantities containing the scatterer's wavenumber, thus the quantities $\Gamma_i$ have to be obtained directly from the $T$ matrix, similarly as was done in references \onlinecite{wu2007effective,zhou2009analytic}. Thus, if we assume that equations \eqref{eq:T0chi0} and \eqref{eq:T2chi2} holds as long as $k_b a<<1$, being $a$ the typical distance between scatterers, we can obtain the frequency-dependent $\Gamma_i$ parameters as
\bea{eq:chiiomega}
\Gamma_0^{11}(\omega)&=&-\frac{4i}{\pi (k_bR_a)^2}T_0^{11}(\omega)\\
\Gamma_0^{12}(\omega)&=&-\frac{4i}{\pi (k_bR_a)^2}T_0^{12}(\omega)\\
\Gamma_2(\omega)&=&-\frac{4i}{\pi (k_bR_a)^2}T_2^{22}(\omega)
\eea
which inserted into equations \eqref{eq:effparams} gives the frequency-dependent parameters $\rho_{eff}(\omega)$, $D_{eff}(\omega)$ and $\nu_{eff}(\omega)$. 
\begin{figure}
\centering
\includegraphics[scale=1]{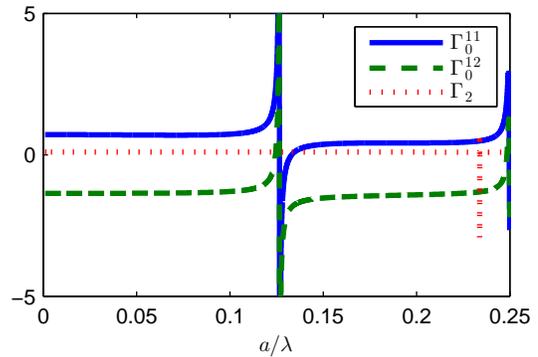}
\caption{\label{fig:chii} Real part of the frequency-dependent parameters $\Gamma_0,\Gamma_1$ and $\Gamma_2$ for rubber inclusions in an aluminium matrix. Only the $\Gamma_0$ and $\Gamma_1$ parameters present
a resonance, while the $\Gamma_2$ parameter is nearly constant along all the frequency range.}
\end{figure}

Figure \ref{fig:chii} shows the three frequency dependent $\Gamma_i$ parameters as a function of frequency
for the case of rubber inclusions in an aluminium plate of thickness $h_b=0.1a$, since the holes and the Lead inclusions does not present low
frequency resonances. It can be seen that only the $\Gamma_0^{11}$ and $\Gamma_0^{12}$ elements present a resonance in a
low enough frequency, that is, for wavelengths in the background such that $\lambda>4a$, where the medium
could be described as an effective medium. A sharp resonance for the $\Gamma_2$ element is found at higher
frequencies near the homogenization limit ($\lambda\approx 4a$), where the description of the effective medium, as will be seen later, has to be taken carefully.

The cut-off wavelength $\lambda>4a$ is chosen in such a way that the wavelength in the background be large enough to see an effective medium, but in the framework of the present theory, this value ensures that we can use the asymptotic form of Bessel functions for small arguments. Also, it can be easily verified that in this frequency range the wavelength inside the scatterer is still larger than the plate's thickness, what ensures that Eq.\eqref{eq:wave} is still a good approximation for the description of flexural waves.

Given that the $\Gamma_0^{11}$ and $\Gamma_0^{12}$ elements appear in the constitutive equations for the effective mass density, rigidity and Poisson's ratio (see equations \eqref{eq:effparams}), the resonance of these elements will affect
all the effective parameters, as will be seen below.
\begin{figure}
\centering
\includegraphics[scale=1]{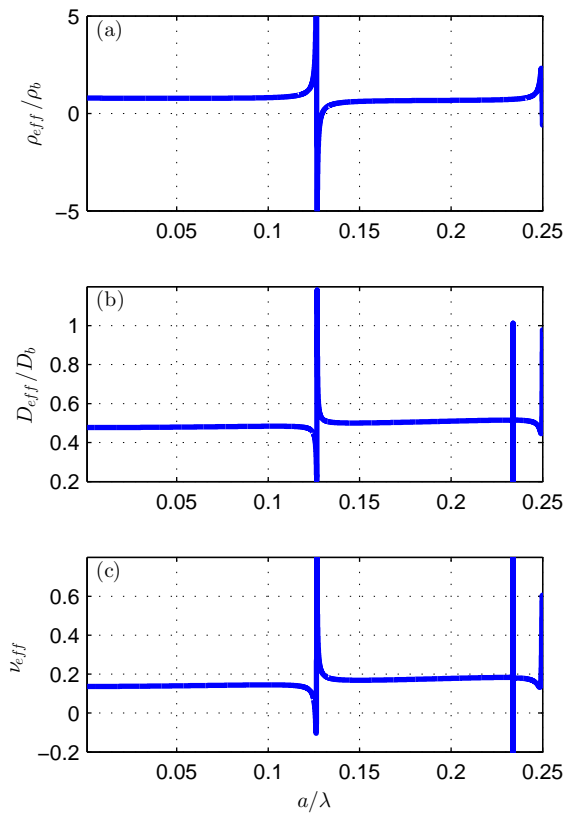}
\caption{\label{fig:effparamsrubber}
 Frequency-dependent effective mass density (a), rigidity (b) and Poisson's ratio (c) for the system described in Fig\ref{fig:chii}. The monopolar resonance creates a resonant behaviour in the three parameters, although only for the mass density is divergent.}
\end{figure}

Figure \ref{fig:effparamsrubber} panel (a) shows the effective mass density for the mentioned system of rubber inclusions. It is clear that the monopolar resonance produces a resonance in the mass density, which allows it to have
positively and negatively divergent values. The monopolar resonance also affects the effective rigidity $D_{eff}$ and Poisson's ratio $\nu_{eff}$, as can be seen from equations \eqref{eq:effparams} and panels (b) and (c) of Figure \ref{fig:effparamsrubber}, however the resonance does not produce any divergent value, since the role in which the $\Gamma_i$ quantities appear is in the form of $\Gamma_0^{11}-\Gamma_0^{12}$, which cancels near the resonance. The resonance of the $\Gamma_2$ coefficient produces a divergent value for both $D_{eff}$ and $\nu_{eff}$, however this resonance is too sharp and occurs too near of the homogenization limit, where the presented theory could be not valid (this limit will be further analysed in Section \ref{sec:numval}).

Rubber inclusions in an elastic matrix can therefore produce a resonant negative
mass density, given that the monopolar resonance appears at low frequencies. However, a good quadrupolar resonator will be required if double negative metamaterials are seek. 

It must be pointed out that expressions \eqref{eq:chiiomega} are valid for any isotropic scatterer, that is, a scatterer with circular symmetry. Therefore, independently of the complexity of the scatterer or the resonator, as long as it can be described by a diagonal $T$ matrix in the low frequency limit, and that no other Lamb waves be excited in the scattering process, the theory can be applied. For instance, the present theory could be applied to systems of spring-mass resonators, attached pillars or spheres, etc.

\section{Numerical Validation}
\label{sec:numval}
The effective medium theory presented in the previous sections has been derived in the low-frequency limit, and assuming that the multiple scattering terms are negligible. Therefore, its validity will be limited to some ``cut-off'' frequency, corresponding to wavelengths in which the field will detect the individual scatterers forming the array, and also it will be limited to some maximum filling fraction in which the multiple scattering elements will be important. This section aims not only to verify the presented theory by full wave multiple scattering simulations, but also gives some approximated values to these limits. Also, subsection \ref{sec:FEM} compares the predicted dispersion relations with the three-dimensional dispersion relation obtained by the finite element method using the commercial software COMSOL Multiphysics.

Multiple scattering theory (MST) allows for the computation of the total field scattered by arbitrarily located inclusions. This method is described for instance in reference \onlinecite{lee2010scattering} for flexural waves, and here it is employed to verify the effective medium theory developed in the previous sections. The verification method consists on the comparison of the scattering properties of a large circular cluster of inclusions with those of an scatterer with the corresponding effective parameters. 

Therefore, according to MST, the total scattered field by a cluster of $N$ inclusions, located at positions $\bm{r}=\bm{R}_\alpha$, with $\alpha=1,2,\ldots,N$, is given by the addition of the scattered field by each inclusion, 
\beq
 W_{sc}=\sum_{\alpha=1}^N W^\alpha_{sc}
\eeq
where
\beq
W_{sc}^\alpha=\sum_{q=-\infty}^\infty\left[B_{q\alpha}^HH_q(k_br_\alpha)+B_{q\alpha}^KK_q(k_br_\alpha)\right]e^{iq\theta_\alpha}
\eeq
and the $B_q^i$ coefficients are the unknowns to be determined. Multiple scattering theory solves for these coefficients by means of a linear system of equations, and the solution is obtained as a function of the frequency of the incident field and the relative position of all the scatterers in the cluster.

In the far field, only the contribution of Hankel functions will be important, since the $K_q$ functions decay exponentially, thus, employing the asymptotic form of Hankel functions, in the far field we have that
\beq
W_{sc}\approx \frac{e^{ik_br}}{\sqrt{r}}F(\theta,k_b)
\eeq
where the far field amplitude $F(\theta,k_b)$ is given by
\beq
F(\theta,k_b)=\frac{e^{-i\pi/4}}{\sqrt{k_b}}\sum_{\alpha=1}^{N}\sum_{q=-\infty}^{\infty}B_{q\alpha}^He^{-ik_bR_\alpha\cos(\theta-\Phi_\alpha)}e^{iq\theta}
\eeq
being $(R_\alpha,\Phi_\alpha)$ the polar coordinates of the $\alpha$-inclusion.

Related with the far field amplitude, and another quantity useful to define the scattering properties of the cluster, we find that the total scattering cross section is given by\cite{NorrisVemula}
\beq
\sigma=\frac{1}{2}\int_0^{2\pi}|F(\theta)|^2d\theta=-2\sqrt{\frac{\pi}{k_b}}Re F(0).
\eeq

In Subsection \ref{sec:ffield} the far field amplitude and the total scattering cross section are used as reference quantities to check the validity of the developed theory. First, in the non-resonant case and later in the resonant one. The following numerical procedure is then followed: A circular cluster of inclusions is taken and its effective parameters are computed. After that, for both the cluster of inclusions and the effective scatterer, the far field amplitude in the forward ($\theta=0$) and backward ($\theta=\pi$) directions and the total scattering cross section are computed and compared for several filling fractions. The comparison is performed in the frequency range that can be considered the homogenization region, corresponding to wavelengths $\lambda>4a$, as was explained before. This numerical experiment gives an idea of the conditions under which these quantities are similar, and then it gives an idea of the conditions under which these systems are indistinguishable, in other words, when the cluster can be described entirely by its effective material.

The reference cluster has been taken as a circular cluster of 151 scatterers arranged in a triangular lattice, as shown in Fig.\ref{fig:schematics}. This cluster has a radius $R_{eff}=6.581a$, being $a$ the lattice constant of the triangular arrangement, and it was employed in references \onlinecite{homoDani1} and \onlinecite{homoDani2} for acoustic waves, showing that it was a big enough cluster to avoid the possible influence of the cluster's boundary in the effective parameters. 

It must be pointed out that, although the effective medium theory has been developed using only the dominant terms of the $T$ matrix, i.e., the multipolar orders $q=0$ and $q=2$, the multiple scattering simulations have been performed using a number of multipoles large enough to ensure convergence, being this number $q=5$ in the present work.

In subsection \ref{sec:mst} some full wave simulations are shown to demonstrate that the field patterns are very similar even in the near field, that is, around the cluster and inside it. Finally, as mentioned before, in subsection \ref{sec:FEM} the theory is validated outside the framework of the flexural wave approximation, that is, by full 3D calculations based on the finite element method.

\subsection{Far-field Analysis}
\label{sec:ffield}
This subsection analyses the far field amplitude along two directions, $\theta=0$, also named the ``forward scattered field'', and $\theta=\pi$, also named the ``back scattered field''. These values are studied also for the three different type of inclusions studied, i.e., holes, rubber and lead inclusions in an aluminium plate. The study of this quantity is made as a function of frequency and for three different filling fractions $f$, being representative values of low ($f=0.04$), mid ($f=0.33$) and high ($f=0.74$) filling fractions.

\begin{figure}
\centering
\includegraphics[scale=1]{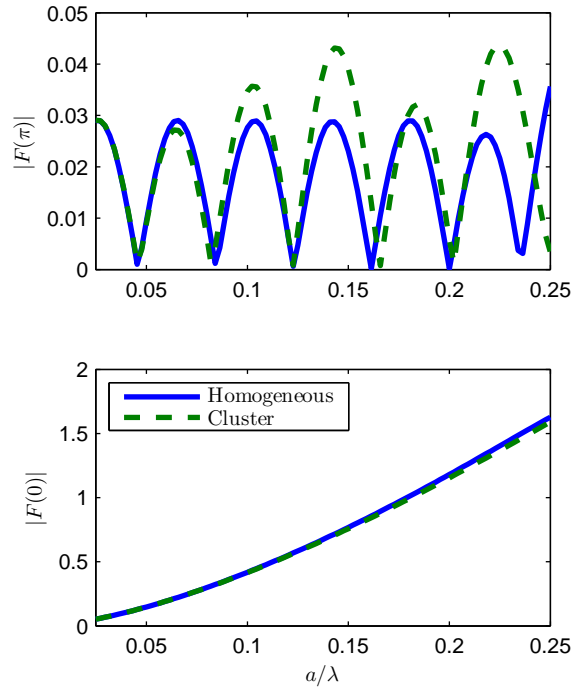}
\caption{\label{fig:sholesR0c1a}(Color online) Far field amplitude in the backward (upper panel) and forward (lower panel) directions as a function of frequency for a circular cluster of 151 holes of radius $R_a=0.1a$ embedded in an triangular lattice of lattice constant $a$ (dashed-green line) compared with that of the effective homogeneous scatterer obtained with the presented homogenization theory}
\end{figure}

Figure \ref{fig:sholesR0c1a} shows the back (upper panel) and the forward (lower panel) scattered fields for the mentioned cluster of holes when the radius of them is $R_a=0.1a$, corresponding to a filling fraction of $f=0.04$, compared with the corresponding effective scatterer whose parameters are given by equations \eqref{eq:effholes}, being $\rho_{eff}=0.96\rho_b$, $D_{eff}=0.92D_b$ and $\nu_{eff}=0.32$. It is shown how small deviations appear in the back scattered field for wavelengths such that $a/\lambda>0.1$, while the forward field patterns are identical in the full frequency region that we could consider the ``homogenization region'', corresponding to wavelengths such that $a/\lambda<0.25$. It is clear that in the very low frequency limit the two quantities are identical.

\begin{figure}
\centering
\includegraphics[scale=1]{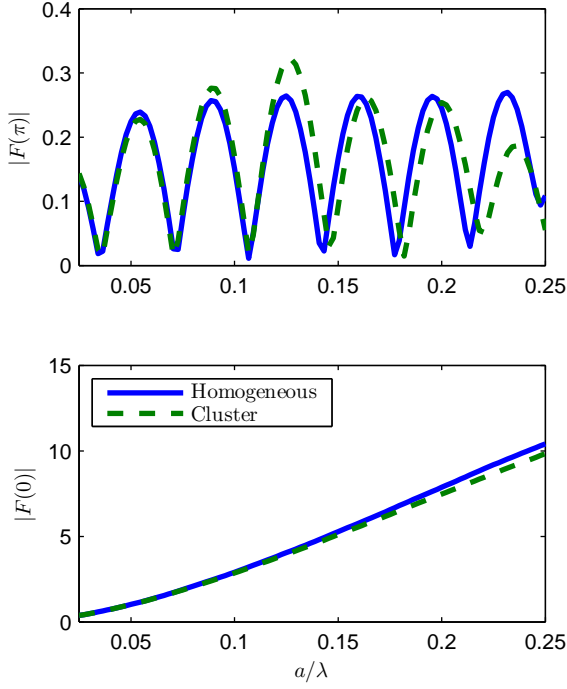}
\caption{\label{fig:sholesR0c3a} (Color online) Same system of Fig. \ref{fig:sholesR0c1a} but being the radius of the holes $R_a=0.3a$.}
\end{figure}

\begin{figure}
\centering
\includegraphics[scale=1]{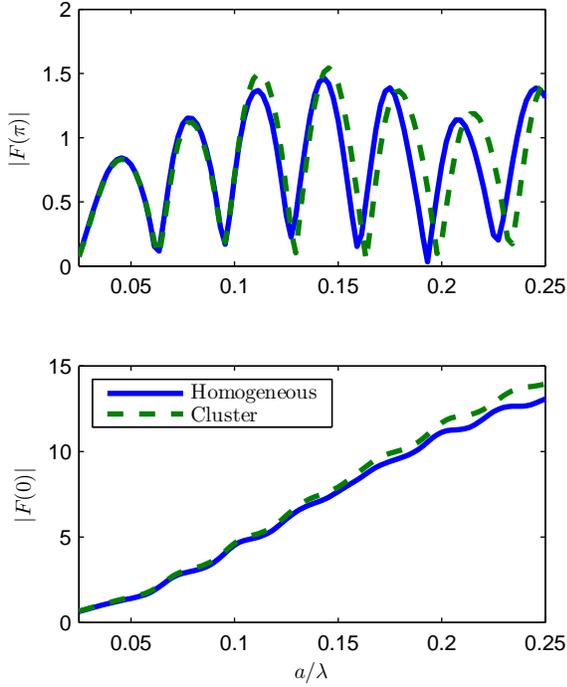}
\caption{\label{fig:sholesR0c45a} (Color online) Same system of Fig. \ref{fig:sholesR0c1a} but being the radius of the holes $R_a=0.45a$.}
\end{figure}

\begin{figure}
\centering
\includegraphics[width=\columnwidth]{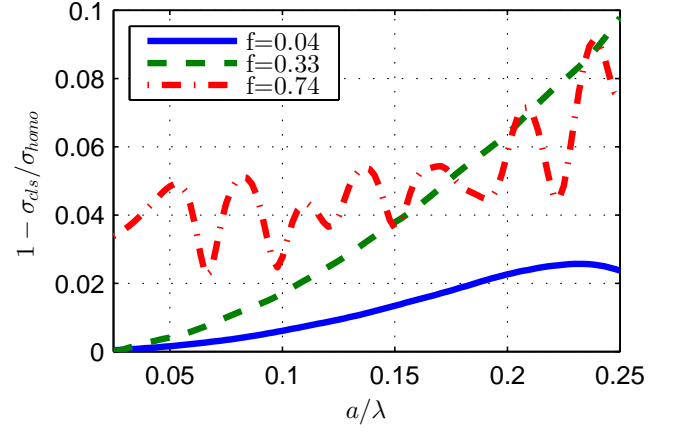}
\caption{\label{fig:sigmarelholes}(Color online) Relative total scattering cross section of the cluster of holes and the corresponding effective parameters. Results are shown for three filling fractions, $f=0.04$(blue continuous line), $f=0.33$(green dashed line) and $0.74$ (red dash-doted line). As can be seen, the differences are higher for higher filling fractions and frequency, but always smaller than 0.1.}
\end{figure}

Similarly, figures \ref{fig:sholesR0c3a} and \ref{fig:sholesR0c45a} show the same quantities but for radius of holes of $R_a=0.3a$ ($f=0.33$) and $R_a=0.45a$ ($f=0.74$), respectively, with effective parameters $\rho_{eff}=0.67\rho_b$, $D_{eff}=0.48D_b$ and $\nu_{eff}=0.14$ for the former and $\rho_{eff}=0.27\rho_b$, $D_{eff}=0.15D_b$ and $\nu_{eff}=-0.03$ for the latter. The negative value for the Poisson ratio in the last case suggest that the theory is not valid for strong scatterers at high filling fraction, as can be deduced from these figures. It is clear that in the two situations the patterns are very similar, however for the high filling fraction cluster the differences are higher as we reach the homogenization limit.
To better visualize these differences, Fig. \ref{fig:sigmarelholes} shows the relative total scattering cross section, defined as
\beq
\sigma_{rel}=\left|\frac{\sigma_{cls}-\sigma_{homo}}{\sigma_{homo}}\right|
\eeq
as a function of frequency and for the three different filling fractions studied. It is clear that even in the high filling fraction situation, the relative differences are never higher than 0.1, which shows that the effective medium approximation is good enough for most applications, although the multiple scattering corrections could be needed if more precision is required, as was done for the design of gradient index lenses or black holes in references \onlinecite{Martin2010,Climente2010,Climente2012}.

\begin{figure}
\centering
\includegraphics[scale=1]{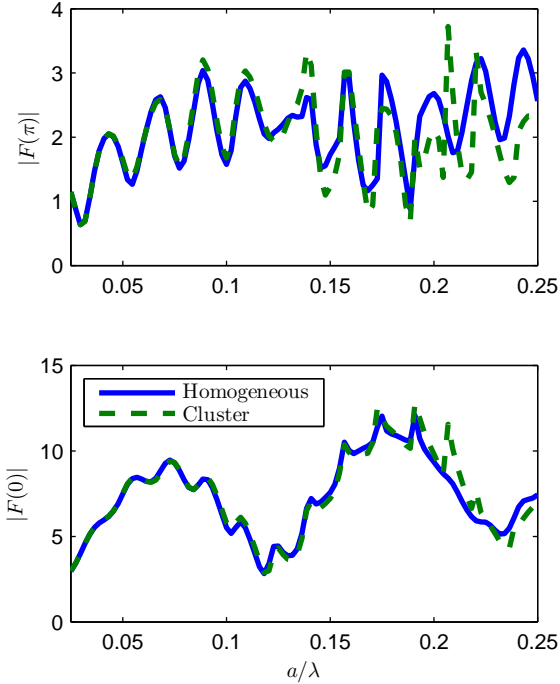}
\caption{\label{fig:sleadR0c45a}(Color online) Same system of Fig.\ref{fig:sholesR0c1a} but with lead inclusions in the aluminium matrix, being the radius of the inclusions $R_a=0.4a$. Even being a situation of high filling fraction, the far field amplitude in the back and forward directions are very similar.}
\end{figure}
\begin{figure}
\centering
\includegraphics[scale=1]{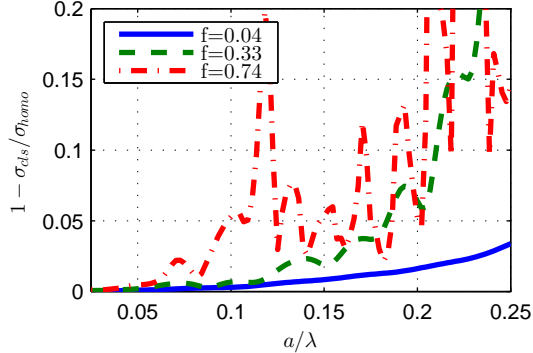}
\caption{\label{fig:sigmarellead} (Color online) Relative total scattering cross section of the cluster of lead inclusions and the corresponding effective parameters. Results are shown for three filling fractions, $f=0.04$ (blue continuous line), $f=0.33$ (green dashed line) and $0.74$ (red dash-doted line). As can be seen,the differences are higher for higher filling fractions and frequency, showing that for the high filling fraction the differences are very important for $a/\lambda>0.1$.}
\end{figure}
A similar analysis has been done for the case of lead inclusions in an aluminium plate, and Fig. \ref{fig:sleadR0c45a} shows the results for the higher filling fraction, corresponding to a radius of inclusions $R_a=0.45a$, being the corresponding effective parameters, given by equations \eqref{eq:effparams}, $\rho_{eff}=3.34\rho_b$, $D_{eff}=0.39D_b$ and $\nu_{eff}=0.32$. The same conclusions as before can be applied, although the relative differences are higher, as can be seen from Fig. \ref{fig:sigmarellead}, which suggest that the multiple scattering corrections could be more important here. However, the relative difference is still smaller than 0.1 in almost all the frequency range. 

When the scatterers are rubber inclusions, the comparison is made with two homogeneous effective scatterers, one employing the frequency-independent theory, whose effective parameters are given by equations \eqref{eq:nonresonant} and the other one as given by the frequency-dependent theory using equations \eqref{eq:chiiomega}. In the following lines, the homogeneous scatterer obtained by means of the frequency-independent theory will be referred as the ``homogeneous scatterer'', while that obtained with the frequency-dependent theory will be referred as the ``metamaterial''.

\begin{figure}
\centering
\includegraphics[scale=1]{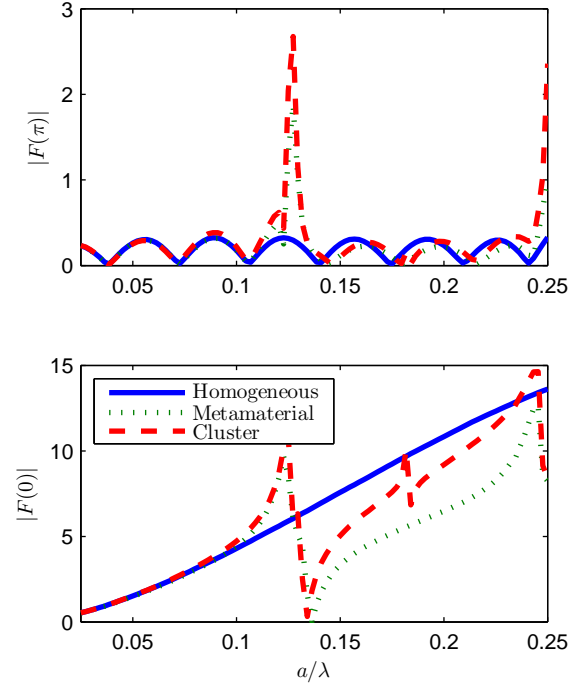}
\caption{\label{fig:srubberR0c3a} (Colour online) Far field amplitude in the back (upper panel) and forward (lower panel) directions for a cluster of 151 rubber inclusions in an aluminium plate (red dashed lines), for a homogeneous scatterer with frequency-independent parameters (blue continuous line) and with a metamaterial with frequency-dependent parameters (green doted line). The radius of the inclusions is $R_a=0.3a$, being the effective parameters of the cluster those depicted in Fig.\ref{fig:effparamsrubber}. It is seen how the frequency-dependent theory predicts better the behaviour of the cluster, although some corrections must be added near the homogenization limit}. 
\end{figure}

Figure \ref{fig:srubberR0c3a} shows the comparison of the back scattered field (upper panel) and the forward field (lower panel) for the cluster of rubber inclusions, the homogeneous scatterer and the metamaterial. The radius of the inclusions is $R_a=0.3a$, thus the effective parameters are the same as those shown in Fig. \ref{fig:effparamsrubber}. Both plots show that the metamaterial and the cluster have more similar scattering properties than the homogeneous scatterer, what shows that the cluster of scatterers is better described by the frequency-dependent theory. However, we can also see that the multiple scattering corrections appear more necessary for frequencies near the homogenization limit $a/\lambda>0.2a$, where the differences are more evident.

Notice that both the metamaterial and the cluster present a resonance near $a/\lambda=0.14$, as can be expected from Fig.\ref{fig:effparamsrubber} and it is due to the monopolar term. Additionally, there is
a resonance in the cluster that is not predicted by the model, and it is given by the dipolar term. However, the effect of this resonance is due to the presence of multiple scattering terms, which should be included in an improved version of the theory, but this is beyond the scope of the present work.

\begin{figure}
\centering
\includegraphics[scale=1]{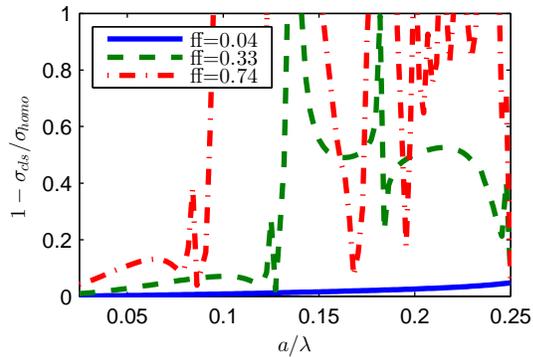}
\caption{\label{fig:sigmarelrubber} (Color online) Relative total scattering cross section for different filling fractions for the cluster of rubber inclusions in an aluminium plate compared with that of the frequency-dependent effective scatterer. It is seen how the differences are important for high frequencies and filling fractions, being therefore necessary the inclusion of the multiple scattering corrections. However, the behaviour of the cluster is properly predicted for low filling fraction and even near the resonance for the filling fraction $f=0.33$.}
\end{figure}

Finally, Fig. \ref{fig:sigmarelrubber}, shows the relative total scattering cross section for three filling fractions, as in the previous sections. It is seen that there is a prefect agreement for the low filling fraction case, $f=0.04$, while we see that the differences are higher for $f=0.33$ and $f=0.74$.
Therefore, the presented results suggest that setting the homogenization limit as wavelengths $\lambda>4a$ is a good approximation for almost all the filling fractions in the non-resonant case, although for high filling fractions this limit should be carefully analysed, and the multiple scattering corrections might be necessary.

The frequency-dependent theory is accurate for low and mid filling fractions, although in this last case is not valid for the full frequency range. The multiple scattering corrections are necessary in this case for a better description of this system, since it is a stronger scattering medium.


\subsection{Near Field Calculations}
\label{sec:mst}
\begin{figure}
\centering
\includegraphics[width=\columnwidth]{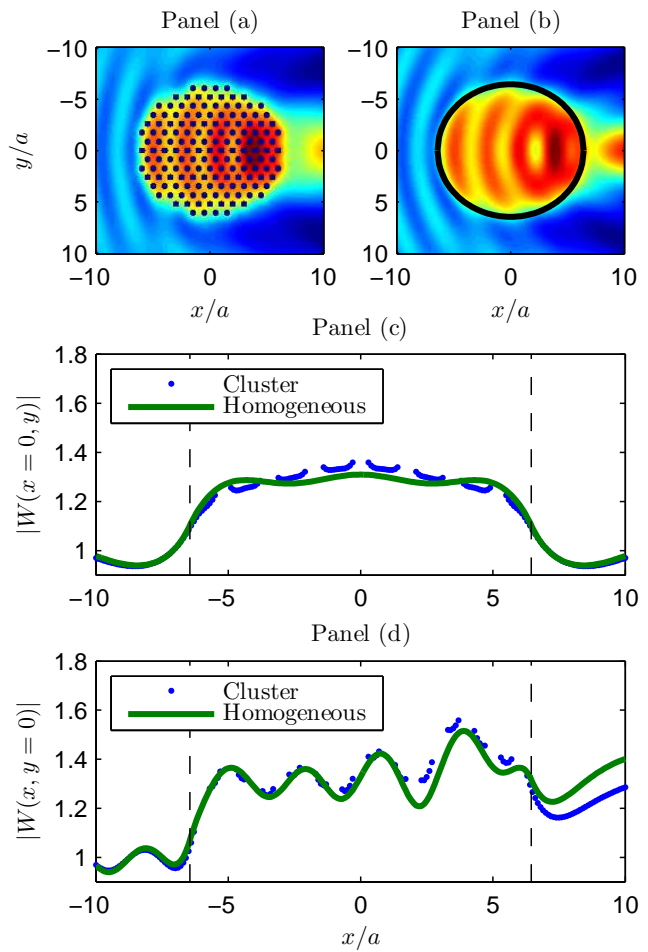}
\caption{\label{fig:mstholes} (a) Multiple scattering simulation of a plane wave interacting with circular cluster of holes in aluminium. (b) Scattering  of the same wave by a single scatterer with the corresponding effective parameters. (c) and (d) Field profiles along the $y$ axis for $x=0$ and along the $x$ axis for $y=0$, respectively, for the cluster (blue dots) and the effective scatterer (green line).}
\end{figure}

In this subsection the field patterns in the near field are shown to validate the theory. Figure \ref{fig:mstholes}, panel (a), shows the field pattern computed by means multiple scattering of the cluster of holes with radius $R_a=0.3a$. A plane wave of wavelength $\lambda=6a$ comes from the left and impinges the cluster, producing the multiple scattering process illustrated. Panel (b) shows the same wave arriving at the corresponding effective homogeneous scatterer. Notice that the field distributions are very similar in the two figures, which shows that the cluster and the effective scatterer behaves in the same way. Panels (c) and (d) of the same figure shows a cut along the $y$ axis for $x=0$ and along the $x$ axis for $y=0$, respectively, where it is possible to see better that the field distributions are very similar not only in the near field but also inside the cluster and the effective scatterer. Notice that there are some discontinuities in the fields computed for the cluster at the positions of the holes. The vertical dashed lines in panels (c) and (d) show the effective radius of the cluster.

\begin{figure}
\centering
\includegraphics[width=\columnwidth]{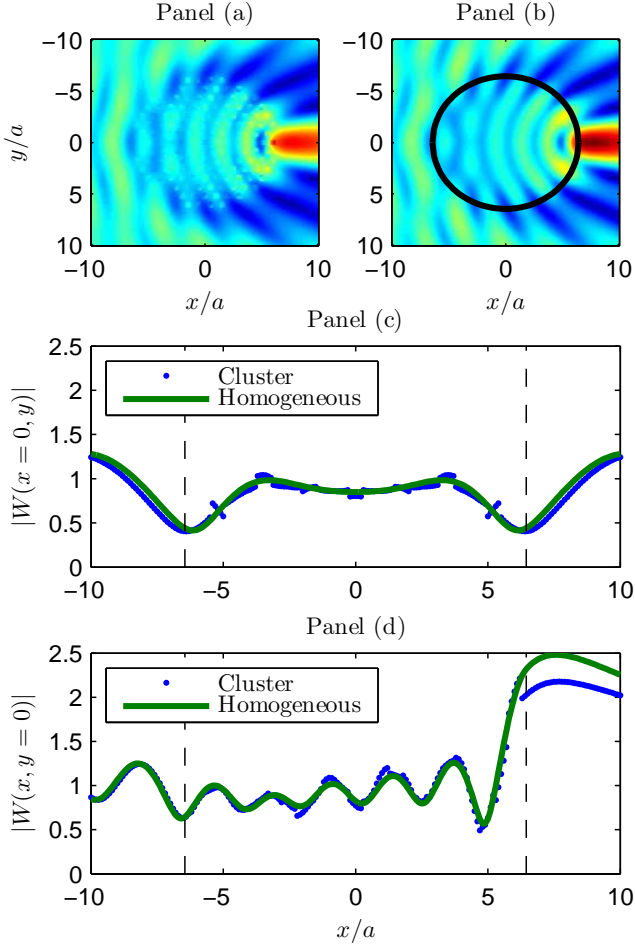}
\caption{\label{fig:mstlead} (a) Multiple scattering simulation of a plane wave interacting with circular cluster of lead inclusions in aluminium. (b) Scattering  of the same wave by a single scatterer with the corresponding effective parameters. (c) and (d) Field profiles along the $y$ axis for $x=0$ and along the $x$ axis for $y=0$, respectively, for the cluster (blue dots) and the effective scatterer (green line).}
\end{figure}

Figure \ref{fig:mstlead} shows the same situation but for a cluster of lead inclusions of the same radius as the holes. The agreement is also obvious here.

\begin{figure}
\centering
\includegraphics[width=\columnwidth]{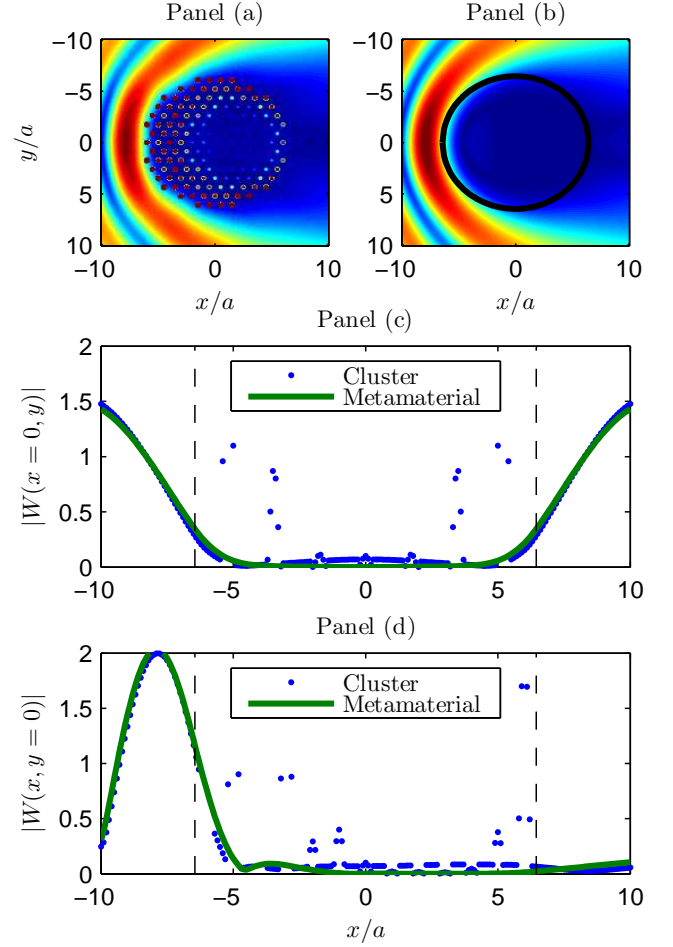}
\caption{\label{fig:mstrubber} (a) Multiple scattering simulation of a plane wave interacting with circular cluster of rubber inclusions in aluminium. (b) Scattering  of the same wave by a single scatterer with the corresponding frequency-dependent effective parameters. (c) and (d) Field profiles along the $y$ axis for $x=0$ and along the $x$ axis for $y=0$, respectively, for the cluster (blue dots) and the effective scatterer (green line).}
\end{figure}

Finally, Fig. \ref{fig:mstrubber} shows the cluster of rubber inclusions, of the same radius as before. The corresponding wavelength is $\lambda=7.89a$ ($a/\lambda=0.127$), where a negative mass density is expected according to Fig. \ref{fig:effparamsrubber}. We see from panel (a) that there is no propagation inside the cluster, since the effective wavenumber is complex and the field is exponentially decaying inside the cluster. In this case, the effective parameters couldn't be obtained by means of equations \eqref{eq:nonresonant}, since they would give us positive parameters and we would observe propagation inside the effective inclusion. Instead, by means of equations \eqref{eq:effparams} and \eqref{eq:chiiomega}, the corresponding effective parameters are $\rho_{eff}=-5.4\rho_b, D_{eff}=1.13D_b$ and $\nu_{eff}=0.64$, which give us a negative mass density and, as we see from panel (b), the field patterns are identical for both the cluster and the effective medium. Panels (c) and (d) shows the corresponding cuts along the main axis, where we see how the field distributions are very similar both inside and outside the cluster. It is remarkable that the exponentially trend of the field inside the cluster is perfectly reproduced by the effective medium theory. There are some peaks in the field distribution of the cluster given to the resonance of the scatterers, which can also be observed in panel (a).

\subsection{3D Dispersion Relations}
\label{sec:FEM}
The previous calculations show the validity of the method within the framework of the flexural wave approximation, valid for wavelengths much larger than
the plate's thickness. In this section we verify that the method is valid within the frequency range of interest by comparing the low-frequency dispersion relation of the three-dimensional plate, computed by the finite element method (FEM), with the expected dispersion relation obtained with the presented theory.
The mentioned calculations have been performed for the two isotropic lattices, the square and the triangular ones, and for two inclusions' radii, $R_a=0.3a$ and $R_a=0.45a$. It must be remarked that depending on the lattice the same radius o the inclusion gives different filling fractions. Notice that the vertical axis of the dispersion relations is the background's wavenumber, related with the frequency by means of Eq.\eqref{eq:kb}.

\begin{figure}
\centering
\includegraphics[width=\columnwidth]{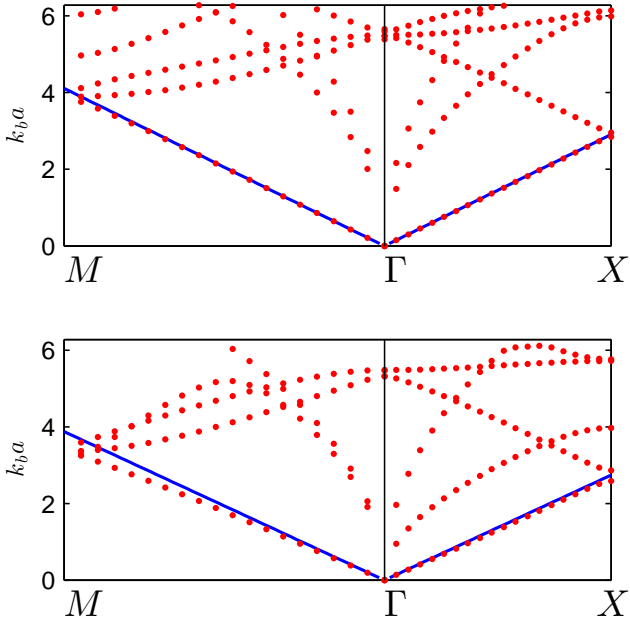}
\caption{\label{fig:SqHoles} Red dots: Three dimensional band structure of a square array of holes of radius $R_a=0.3a$ (upper panel) and $R_a=0.45a$ (lower panel) in an aluminium plate of thickness $h_b=0.1a$ computed by FEM. Blue line: Linear dispersion relation obtained with the effective medium theory of the present work. }
\end{figure}

Figure \ref{fig:SqHoles} shows the dispersion relation of an aluminium plate of thickness $h_b=0.1a$ with a square arrangement of holes computed by FEM (red dots) compared with the linear dispersion relation predicted by the model (blue line). Results are shown for a radius of the holes of $R_a=0.3a$ (upper panel) and $R_a=0.45a$ (lower panel) and for the two symmetry directions of the lattice. It is obvious that the dispersion relations are very similar for the two filling fractions. As expected, the medium is isotropic in the low frequency limit.

\begin{figure}
\centering
\includegraphics[width=\columnwidth]{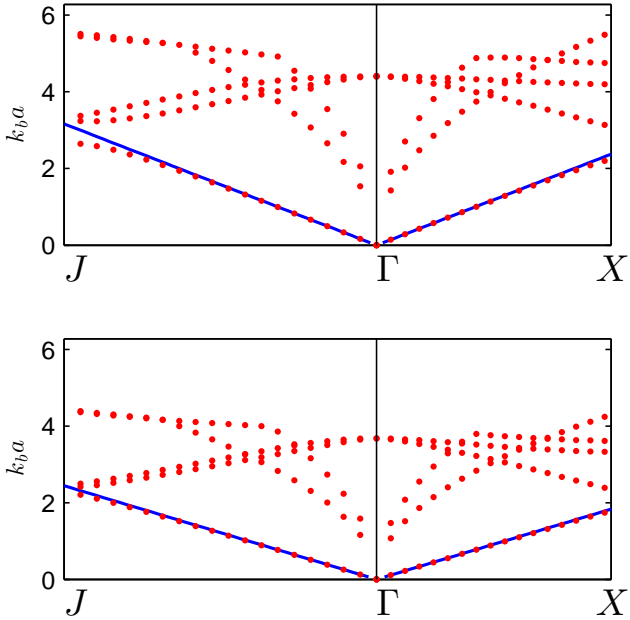}
\caption{\label{fig:TrLead} Red dots: Three dimensional band structure of a triangular array of lead inclusions of radius $R_a=0.3a$ (upper panel) and $R_a=0.45a$ (lower panel) in an aluminium plate of thickness $h_b=0.1a$ computed by FEM. Blue line: Linear dispersion relation obtained with the effective medium theory of the present work.}
\end{figure}

Figure \ref{fig:TrLead} shows the same situation but for a triangular lattice of lead inclusions, with the same radii of inclusions. Again, there is a good agreement between the full band structure computed by FEM and the presented theory. 

\begin{figure}
\centering
\includegraphics[width=\columnwidth]{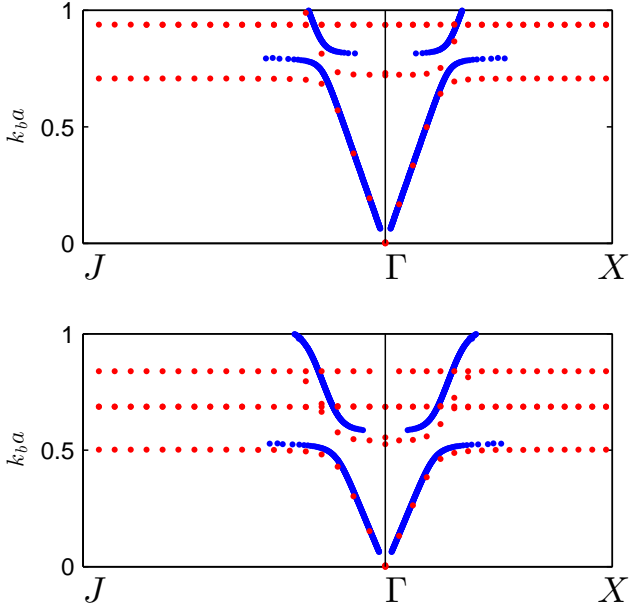}
\caption{\label{fig:TrRubber} Red dots: Three dimensional band structure of a triangular array of rubber inclusions of radius $R_a=0.3a$ (upper panel) and $R_a=0.45a$ (lower panel) in an aluminium plate of thickness $h_b=0.1a$ computed by FEM. Blue line: Linear dispersion relation obtained with the effective medium theory of the present work.}
\end{figure}

Finally, Fig.\ref{fig:TrRubber} shows the comparison of the dispersion curves for the triangular lattice of rubber inclusions and same radii as before. In this case the dispersion relation of the effective medium has been obtained by means of the frequency-dependent theory. It is clear that in the very-low frequency limit the slopes are identical for both filling fractions, and that in both situations there is an opening of the band gap. However, two disagreements are found in these curves. First, more resonances appear in the dispersion relation that are not predicted by the model, second, the position of the resonance is better predicted for the high filling fraction lattice. These disagreements have different origin, as explained below.

The existence of the additional resonances was predicted in the previous sections, by means of the far field amplitude, and it is due to the existence of multiple scattering terms that should be included in the theory. Notice however that these resonances are very sharp, and they are likely to disappear in a hypothetical experiment due to losses in real materials.

The disagreement in the position of the peak in the low filling fraction is given to the accuracy of the flexural wave approximation. Effectively, this approximation is valid for wavelengths larger than the plate's thickness, and for the radius of inclusions of $R_a=0.45a$ the resonance appears around $k_b a\approx 0.5$, which inside rubber gives a wavenumber of $k_a a\approx 6.7$, with a wavelength $\lambda\approx a \approx 10h_b$, so that the wavelength inside rubber is nearly 10 times the thickness of the plate, what makes the flexural wave approximation still valid. However, for the radius of $R_a=0.3a$, this resonance appears around $k_b a\approx	0.7$, being the wavenumber in rubber $k_a a\approx 10$, with a wavelength $\lambda\approx 0.6a\approx 6h_b$, which is still a wavelength larger than the plate's thickness but that could suggest that higher order plates theory may be needed for these frequencies for a better description, although a good approximation is found here.

\section{Summary}
\label{sec:summary}
In summary, an effective medium description for the propagation of flexural waves in thin elastic plates with isotropic arrangements of inclusions or resonators has been presented. The theory is based on the scattering
properties of these inclusions, which are compared with that of an effective scatterer and, from the low frequency limit of the expressions found for the scattering coefficients, closed form expressions for the effective elastic parameters of the arrangement are obtained. 

The theory is valid not only in the quasi-static limit, but also for frequency-dependent effective parameters, situation that happens in the case of having a long wavelength in the background but not inside the scatterers.

For the resonant medium it is found that the negative mass density is obtained from a monopolar resonance, unlike acoustic or bulk elastic waves where this extraordinary behaviour is due to a dipolar resonance. Additionally, the resonant behaviour of both the rigidity and the Poisson's ratio is given to both the monopolar and the quadrupolar resonance, being the latter the most important contribution.

It has been found that, surprisingly, the dipolar term plays no role in the effective  parameters in the present approximation, valid for low and mid filling fractions, however it is expected that including the full multiple-scattering terms in the equations the role of the dipolar term be clarified, but always as a multiple scattering correction. The challenge for this type of metamaterials the finding of a good quadrupolar resonator, in order to obtain doubly negative metamaterials.

Multiple scattering theory was used to numerically verify the presented theory, by the comparison of the scattering properties of a cluster of inclusions with those of the corresponding effective homogeneous scatterer. This comparison was performed for different filling fractions and in the frequency range corresponding to the homogenization limit, defined as wavelengths such that $\lambda>4a$. It has been found that the theory is valid for low and mid filling fractions in practically all the homogenization region, although the multiple scattering corrections must be added for high filling fractions and for strong resonant scatterers. 

Finally, the dispersion relation of the three-dimensional plate, computed by the finite element method using the commercial software COMSOL Multiphysics, is obtained for several lattices, filling fractions and materials, and compared at low frequencies with that predicted by the present method, showing a very good agreement between the two calculations.

This work opens the door for the efficient design of simple, doubly and triply resonant metamaterials for flexural waves in thin plates, and can be the basis for its generalization to all type of Lamb waves, with promising applications in the control of vibrations in all scales.

\acknowledgments{Work supported by the Agence Nationale de la Recherche and Direction G\'en\'erale de l'Armement under the project Metactif, grant ANR-11-ASTR-015.}
\appendix
\section{Closed form expression for the $T$ matrix}
\label{sec:XY}
The expressions for the $\bm{X_q^i}$ and $\bm{Y_q^i}$ matrices are required to obtain the $T$ matrix of a circular inhomogeneity, and they are obtained after applying boundary conditions at $r=R_a$, being
\bea{eq:X0Y0}
X_0&=&
\left(\begin{array}{cc}
J_q(k_b R_a) & I_q(k_bR_a)\\
k_bJ_q'(k_bR_a) & k_bI_q'(k_bR_a)
\end{array}\right) \\
Y_0&=&
\left(\begin{array}{cc}
S_q^J(k_b R_a) & S_q^I(k_bR_a)\\
T_q^J(k_b R_a) & T_q^I(k_bR_a)
\end{array}\right) 
\eea

\bea{eq:XaYa}
X_a&=&
\left(\begin{array}{cc}
J_q(k_a R_a) & I_q(k_aR_a)\\
k_aJ_q'(k_aR_a) & k_aI_q'(k_aR_a)
\end{array}\right)
\\ 
Y_a&=&
\left(\begin{array}{cc}
S_q^J(k_a R_a) & S_q^I(k_aR_a)\\
T_q^J(k_a R_a) & T_q^I(k_aR_a)
\end{array}\right)
\eea

\bea{eq:XscYsc}
X_{sc}&=&
\left(\begin{array}{cc}
H_q(k_b R_a) & K_q(k_bR_a)\\
k_bH_q'(k_bR_a) & k_bK_q'(k_bR_a)
\end{array}\right) \\
Y_{sc}&=&
\left(\begin{array}{cc}
S_q^H(k_b R_a) & S_q^K(k_bR_a)\\
T_q^H(k_bR_a) & T_q^K(k_bR_a)
\end{array}\right)
\eea
where
\begin{widetext} 
\bea{ST}
S_q^X(kr)&=&D\left[q^2(1-\nu)\mp k^2r^2\right]X_q(kr)-D(1-\nu)krX_q'(kr)\\
T_q^X(kr)&=&Dq^2(1-\nu)X_q(kr)-D\left[q^2(1-\nu)\pm k^2r^2\right]krX_q'(kr)
\eea
\end{widetext}
where the upper sign applies when $X$ is $J_q,H_q$ and the lower sign when $X$ is $I_q,K_q$.

The $T$ matrix is defined by means of equation \eqref{eq:BqAq} as
\begin{multline}
\label{eq:Tq}
T_q=-\left(\bm{Y}^{sc}_q-\bm{Y}^a_q(\bm{X}^a_q)^{-1}\bm{X}^{sc}_q\right)^{-1}\times\\
\left(\bm{Y}^0_q-\bm{Y}^a_q(\bm{X}^a_q)^{-1}\bm{X}^0_q\right).
\end{multline}
The low frequency terms of the above expression can be obtained after some manipulation. First, the
inclusion's terms can be developed, being
\eqlabel{eq:YaXam1}{
\bm{Y}^a_q(\bm{X}^a_q)^{-1}=\frac{1}{\Delta}\Mat{a_{11}}{a_{12}}{a_{21}}{a_{22}}
}
where, with $x=k_aR_a$,
\bea{eq:aij}
a_{11}&=&k_a(I_q'(x)S_q^J(x)-J_q'(x)S_q^I(x))\\
a_{12}&=&J_q(x)S_q^I(x)-I_q(x)S_q^J(x)\\
a_{21}&=&k_a(I_q'(x)T_q^J(x)-J_q'(x)T_q^I(x))\\
a_{21}&=&J_q(x)T_q^I(x)-I_q(x)T_q^J(x)\\
\Delta&=&k_a(J_q(x)I_q'(x)-J_q'(x)I_q(x))
\eea
Then we have
\begin{widetext}
\begin{multline}
\label{eq:Hq}
\bm{Y}^{sc}_q-\bm{Y}^a_q(\bm{X}^a_q)^{-1}\bm{X}^{sc}_q=\\
\Mat{S_q^H(k_bR_a)-H_q(k_bR_a)a_{11}/\Delta-k_bH_q'(k_bR_a)a_{12}/\Delta}
{S_q^K(k_bR_a)-K_q(k_bR_a)a_{11}/\Delta-k_bK_q'(k_bR_a)a_{12}/\Delta}
{T_q^H(k_bR_a)-H_q(k_bR_a)a_{21}/\Delta-k_bH_q'(k_bR_a)a_{22}/\Delta}
{T_q^K(k_bR_a)-K_q(k_bR_a)a_{21}/\Delta-k_bK_q'(k_bR_a)a_{22}/\Delta}
\end{multline}
and
\begin{multline}
\label{eq:Jq}
\bm{Y}^{0}_q-\bm{Y}^a_q(\bm{X}^a_q)^{-1}\bm{X}^{0}_q=\\
\Mat{S_q^J(k_bR_a)-J_q(k_bR_a)a_{11}/\Delta-k_bJ_q'(k_bR_a)a_{12}/\Delta}
{S_q^I(k_bR_a)-I_q(k_bR_a)a_{11}/\Delta-k_bI_q'(k_bR_a)a_{12}/\Delta}
{T_q^J(k_bR_a)-J_q(k_bR_a)a_{21}/\Delta-k_bJ_q'(k_bR_a)a_{22}/\Delta}
{T_q^I(k_bR_a)-I_q(k_bR_a)a_{21}/\Delta-k_bI_q'(k_bR_a)a_{22}/\Delta}.
\end{multline}
\end{widetext}

The above expressions are now suitable for their analysis in the low frequency limit, since essentially they contains linear combinations of Bessel, Hankel and modified Bessel functions. The details are tough, but the results given by Eqs.\eqref{eq:chii} have been verified both analytically and with the help of a symbolic math computer software. Then, it has been found that, in the low frequency limit, the dominant terms of the $T$ matrix are the $q=0$ and the $q=2$, since they go to zero as $k_b^2$, while the $q=1$ goes as $k_b^2$. The other elements goes as $k_b^{2q-2}$.


\begin{thebibliography}{51}%
\makeatletter
\providecommand \@ifxundefined [1]{%
 \@ifx{#1\undefined}
}%
\providecommand \@ifnum [1]{%
 \ifnum #1\expandafter \@firstoftwo
 \else \expandafter \@secondoftwo
 \fi
}%
\providecommand \@ifx [1]{%
 \ifx #1\expandafter \@firstoftwo
 \else \expandafter \@secondoftwo
 \fi
}%
\providecommand \natexlab [1]{#1}%
\providecommand \enquote  [1]{``#1''}%
\providecommand \bibnamefont  [1]{#1}%
\providecommand \bibfnamefont [1]{#1}%
\providecommand \citenamefont [1]{#1}%
\providecommand \href@noop [0]{\@secondoftwo}%
\providecommand \href [0]{\begingroup \@sanitize@url \@href}%
\providecommand \@href[1]{\@@startlink{#1}\@@href}%
\providecommand \@@href[1]{\endgroup#1\@@endlink}%
\providecommand \@sanitize@url [0]{\catcode `\\12\catcode `\$12\catcode
  `\&12\catcode `\#12\catcode `\^12\catcode `\_12\catcode `\%12\relax}%
\providecommand \@@startlink[1]{}%
\providecommand \@@endlink[0]{}%
\providecommand \url  [0]{\begingroup\@sanitize@url \@url }%
\providecommand \@url [1]{\endgroup\@href {#1}{\urlprefix }}%
\providecommand \urlprefix  [0]{URL }%
\providecommand \Eprint [0]{\href }%
\providecommand \doibase [0]{http://dx.doi.org/}%
\providecommand \selectlanguage [0]{\@gobble}%
\providecommand \bibinfo  [0]{\@secondoftwo}%
\providecommand \bibfield  [0]{\@secondoftwo}%
\providecommand \translation [1]{[#1]}%
\providecommand \BibitemOpen [0]{}%
\providecommand \bibitemStop [0]{}%
\providecommand \bibitemNoStop [0]{.\EOS\space}%
\providecommand \EOS [0]{\spacefactor3000\relax}%
\providecommand \BibitemShut  [1]{\csname bibitem#1\endcsname}%
\let\auto@bib@innerbib\@empty
\bibitem [{\citenamefont {Smith}\ \emph {et~al.}(2004)\citenamefont {Smith},
  \citenamefont {Pendry},\ and\ \citenamefont
  {Wiltshire}}]{smith2004metamaterials}%
  \BibitemOpen
  \bibfield  {author} {\bibinfo {author} {\bibfnamefont {D.}~\bibnamefont
  {Smith}}, \bibinfo {author} {\bibfnamefont {J.}~\bibnamefont {Pendry}}, \
  and\ \bibinfo {author} {\bibfnamefont {M.}~\bibnamefont {Wiltshire}},\
  }\href@noop {} {\bibfield  {journal} {\bibinfo  {journal} {Science}\ }\textbf
  {\bibinfo {volume} {305}},\ \bibinfo {pages} {788} (\bibinfo {year}
  {2004})}\BibitemShut {NoStop}%
\bibitem [{\citenamefont {Pendry}\ \emph {et~al.}(2006)\citenamefont {Pendry},
  \citenamefont {Schurig},\ and\ \citenamefont
  {Smith}}]{pendry2006controlling}%
  \BibitemOpen
  \bibfield  {author} {\bibinfo {author} {\bibfnamefont {J.~B.}\ \bibnamefont
  {Pendry}}, \bibinfo {author} {\bibfnamefont {D.}~\bibnamefont {Schurig}}, \
  and\ \bibinfo {author} {\bibfnamefont {D.~R.}\ \bibnamefont {Smith}},\
  }\href@noop {} {\bibfield  {journal} {\bibinfo  {journal} {Science}\ }\textbf
  {\bibinfo {volume} {312}},\ \bibinfo {pages} {1780} (\bibinfo {year}
  {2006})}\BibitemShut {NoStop}%
\bibitem [{\citenamefont {Li}\ and\ \citenamefont {Chan}(2004)}]{li2004double}%
  \BibitemOpen
  \bibfield  {author} {\bibinfo {author} {\bibfnamefont {J.}~\bibnamefont
  {Li}}\ and\ \bibinfo {author} {\bibfnamefont {C.}~\bibnamefont {Chan}},\
  }\href@noop {} {\bibfield  {journal} {\bibinfo  {journal} {Physical Review
  E}\ }\textbf {\bibinfo {volume} {70}},\ \bibinfo {pages} {055602} (\bibinfo
  {year} {2004})}\BibitemShut {NoStop}%
\bibitem [{\citenamefont {Cummer}\ and\ \citenamefont
  {Schurig}(2007)}]{cummer2007one}%
  \BibitemOpen
  \bibfield  {author} {\bibinfo {author} {\bibfnamefont {S.~A.}\ \bibnamefont
  {Cummer}}\ and\ \bibinfo {author} {\bibfnamefont {D.}~\bibnamefont
  {Schurig}},\ }\href@noop {} {\bibfield  {journal} {\bibinfo  {journal} {New
  Journal of Physics}\ }\textbf {\bibinfo {volume} {9}},\ \bibinfo {pages} {45}
  (\bibinfo {year} {2007})}\BibitemShut {NoStop}%
\bibitem [{\citenamefont {Liu}\ \emph {et~al.}(2000)\citenamefont {Liu},
  \citenamefont {Zhang}, \citenamefont {Mao}, \citenamefont {Zhu},
  \citenamefont {Yang}, \citenamefont {Chan},\ and\ \citenamefont
  {Sheng}}]{liu2000locally}%
  \BibitemOpen
  \bibfield  {author} {\bibinfo {author} {\bibfnamefont {Z.}~\bibnamefont
  {Liu}}, \bibinfo {author} {\bibfnamefont {X.}~\bibnamefont {Zhang}}, \bibinfo
  {author} {\bibfnamefont {Y.}~\bibnamefont {Mao}}, \bibinfo {author}
  {\bibfnamefont {Y.}~\bibnamefont {Zhu}}, \bibinfo {author} {\bibfnamefont
  {Z.}~\bibnamefont {Yang}}, \bibinfo {author} {\bibfnamefont {C.}~\bibnamefont
  {Chan}}, \ and\ \bibinfo {author} {\bibfnamefont {P.}~\bibnamefont {Sheng}},\
  }\href@noop {} {\bibfield  {journal} {\bibinfo  {journal} {Science}\ }\textbf
  {\bibinfo {volume} {289}},\ \bibinfo {pages} {1734} (\bibinfo {year}
  {2000})}\BibitemShut {NoStop}%
\bibitem [{\citenamefont {Milton}\ \emph {et~al.}(2006)\citenamefont {Milton},
  \citenamefont {Briane},\ and\ \citenamefont {Willis}}]{milton2006cloaking}%
  \BibitemOpen
  \bibfield  {author} {\bibinfo {author} {\bibfnamefont {G.~W.}\ \bibnamefont
  {Milton}}, \bibinfo {author} {\bibfnamefont {M.}~\bibnamefont {Briane}}, \
  and\ \bibinfo {author} {\bibfnamefont {J.~R.}\ \bibnamefont {Willis}},\
  }\href@noop {} {\bibfield  {journal} {\bibinfo  {journal} {New Journal of
  Physics}\ }\textbf {\bibinfo {volume} {8}},\ \bibinfo {pages} {248} (\bibinfo
  {year} {2006})}\BibitemShut {NoStop}%
\bibitem [{\citenamefont {Schurig}\ \emph {et~al.}(2006)\citenamefont
  {Schurig}, \citenamefont {Mock},\ and\ \citenamefont
  {Smith}}]{schurig2006electric}%
  \BibitemOpen
  \bibfield  {author} {\bibinfo {author} {\bibfnamefont {D.}~\bibnamefont
  {Schurig}}, \bibinfo {author} {\bibfnamefont {J.}~\bibnamefont {Mock}}, \
  and\ \bibinfo {author} {\bibfnamefont {D.}~\bibnamefont {Smith}},\
  }\href@noop {} {\bibfield  {journal} {\bibinfo  {journal} {Applied Physics
  Letters}\ }\textbf {\bibinfo {volume} {88}},\ \bibinfo {pages} {041109}
  (\bibinfo {year} {2006})}\BibitemShut {NoStop}%
\bibitem [{\citenamefont {Torrent}\ and\ \citenamefont
  {S{\'a}nchez-Dehesa}(2008)}]{torrent2008anisotropic}%
  \BibitemOpen
  \bibfield  {author} {\bibinfo {author} {\bibfnamefont {D.}~\bibnamefont
  {Torrent}}\ and\ \bibinfo {author} {\bibfnamefont {J.}~\bibnamefont
  {S{\'a}nchez-Dehesa}},\ }\href@noop {} {\bibfield  {journal} {\bibinfo
  {journal} {New journal of physics}\ }\textbf {\bibinfo {volume} {10}},\
  \bibinfo {pages} {023004} (\bibinfo {year} {2008})}\BibitemShut {NoStop}%
\bibitem [{\citenamefont {Fang}\ \emph {et~al.}(2006)\citenamefont {Fang},
  \citenamefont {Xi}, \citenamefont {Xu}, \citenamefont {Ambati}, \citenamefont
  {Srituravanich}, \citenamefont {Sun},\ and\ \citenamefont
  {Zhang}}]{fang2006ultrasonic}%
  \BibitemOpen
  \bibfield  {author} {\bibinfo {author} {\bibfnamefont {N.}~\bibnamefont
  {Fang}}, \bibinfo {author} {\bibfnamefont {D.}~\bibnamefont {Xi}}, \bibinfo
  {author} {\bibfnamefont {J.}~\bibnamefont {Xu}}, \bibinfo {author}
  {\bibfnamefont {M.}~\bibnamefont {Ambati}}, \bibinfo {author} {\bibfnamefont
  {W.}~\bibnamefont {Srituravanich}}, \bibinfo {author} {\bibfnamefont
  {C.}~\bibnamefont {Sun}}, \ and\ \bibinfo {author} {\bibfnamefont
  {X.}~\bibnamefont {Zhang}},\ }\href@noop {} {\bibfield  {journal} {\bibinfo
  {journal} {Nature materials}\ }\textbf {\bibinfo {volume} {5}},\ \bibinfo
  {pages} {452} (\bibinfo {year} {2006})}\BibitemShut {NoStop}%
\bibitem [{\citenamefont {Torrent}\ and\ \citenamefont
  {S{\'a}nchez-Dehesa}(2009)}]{torrent2009radial}%
  \BibitemOpen
  \bibfield  {author} {\bibinfo {author} {\bibfnamefont {D.}~\bibnamefont
  {Torrent}}\ and\ \bibinfo {author} {\bibfnamefont {J.}~\bibnamefont
  {S{\'a}nchez-Dehesa}},\ }\href@noop {} {\bibfield  {journal} {\bibinfo
  {journal} {Physical review letters}\ }\textbf {\bibinfo {volume} {103}},\
  \bibinfo {pages} {064301} (\bibinfo {year} {2009})}\BibitemShut {NoStop}%
\bibitem [{\citenamefont {Torrent}\ and\ \citenamefont
  {S{\'a}nchez-Dehesa}(2010)}]{torrent2010acoustic}%
  \BibitemOpen
  \bibfield  {author} {\bibinfo {author} {\bibfnamefont {D.}~\bibnamefont
  {Torrent}}\ and\ \bibinfo {author} {\bibfnamefont {J.}~\bibnamefont
  {S{\'a}nchez-Dehesa}},\ }\href@noop {} {\bibfield  {journal} {\bibinfo
  {journal} {New Journal of Physics}\ }\textbf {\bibinfo {volume} {12}},\
  \bibinfo {pages} {073034} (\bibinfo {year} {2010})}\BibitemShut {NoStop}%
\bibitem [{\citenamefont {Christensen}\ \emph {et~al.}(2007)\citenamefont
  {Christensen}, \citenamefont {Fernandez-Dominguez}, \citenamefont
  {de~Leon-Perez}, \citenamefont {Martin-Moreno},\ and\ \citenamefont
  {Garcia-Vidal}}]{christensen2007collimation}%
  \BibitemOpen
  \bibfield  {author} {\bibinfo {author} {\bibfnamefont {J.}~\bibnamefont
  {Christensen}}, \bibinfo {author} {\bibfnamefont {A.}~\bibnamefont
  {Fernandez-Dominguez}}, \bibinfo {author} {\bibfnamefont {F.}~\bibnamefont
  {de~Leon-Perez}}, \bibinfo {author} {\bibfnamefont {L.}~\bibnamefont
  {Martin-Moreno}}, \ and\ \bibinfo {author} {\bibfnamefont {F.}~\bibnamefont
  {Garcia-Vidal}},\ }\href@noop {} {\bibfield  {journal} {\bibinfo  {journal}
  {Nature Physics}\ }\textbf {\bibinfo {volume} {3}},\ \bibinfo {pages} {851}
  (\bibinfo {year} {2007})}\BibitemShut {NoStop}%
\bibitem [{\citenamefont {Torrent}\ and\ \citenamefont
  {S{\'a}nchez-Dehesa}(2012)}]{torrent2012acoustic}%
  \BibitemOpen
  \bibfield  {author} {\bibinfo {author} {\bibfnamefont {D.}~\bibnamefont
  {Torrent}}\ and\ \bibinfo {author} {\bibfnamefont {J.}~\bibnamefont
  {S{\'a}nchez-Dehesa}},\ }\href@noop {} {\bibfield  {journal} {\bibinfo
  {journal} {Physical review letters}\ }\textbf {\bibinfo {volume} {108}},\
  \bibinfo {pages} {174301} (\bibinfo {year} {2012})}\BibitemShut {NoStop}%
\bibitem [{\citenamefont {Farhat}\ \emph
  {et~al.}(2009{\natexlab{a}})\citenamefont {Farhat}, \citenamefont {Guenneau},
  \citenamefont {Enoch},\ and\ \citenamefont {Movchan}}]{Farhat2009}%
  \BibitemOpen
  \bibfield  {author} {\bibinfo {author} {\bibfnamefont {M.}~\bibnamefont
  {Farhat}}, \bibinfo {author} {\bibfnamefont {S.}~\bibnamefont {Guenneau}},
  \bibinfo {author} {\bibfnamefont {S.}~\bibnamefont {Enoch}}, \ and\ \bibinfo
  {author} {\bibfnamefont {A.}~\bibnamefont {Movchan}},\ }\href {\doibase
  10.1103/PhysRevB.79.033102} {\bibfield  {journal} {\bibinfo  {journal}
  {Physical Review B}\ }\textbf {\bibinfo {volume} {79}},\ \bibinfo {pages}
  {033102} (\bibinfo {year} {2009}{\natexlab{a}})}\BibitemShut {NoStop}%
\bibitem [{\citenamefont {Farhat}\ \emph
  {et~al.}(2009{\natexlab{b}})\citenamefont {Farhat}, \citenamefont
  {Guenneau},\ and\ \citenamefont {Enoch}}]{Farhat2009a}%
  \BibitemOpen
  \bibfield  {author} {\bibinfo {author} {\bibfnamefont {M.}~\bibnamefont
  {Farhat}}, \bibinfo {author} {\bibfnamefont {S.}~\bibnamefont {Guenneau}}, \
  and\ \bibinfo {author} {\bibfnamefont {S.}~\bibnamefont {Enoch}},\ }\href
  {\doibase 10.1103/PhysRevLett.103.024301} {\bibfield  {journal} {\bibinfo
  {journal} {Physical Review Letters}\ }\textbf {\bibinfo {volume} {103}},\
  \bibinfo {pages} {1} (\bibinfo {year} {2009}{\natexlab{b}})}\BibitemShut
  {NoStop}%
\bibitem [{\citenamefont {Stenger}\ \emph {et~al.}(2012)\citenamefont
  {Stenger}, \citenamefont {Wilhelm},\ and\ \citenamefont
  {Wegener}}]{Stenger2012}%
  \BibitemOpen
  \bibfield  {author} {\bibinfo {author} {\bibfnamefont {N.}~\bibnamefont
  {Stenger}}, \bibinfo {author} {\bibfnamefont {M.}~\bibnamefont {Wilhelm}}, \
  and\ \bibinfo {author} {\bibfnamefont {M.}~\bibnamefont {Wegener}},\ }\href
  {\doibase 10.1103/PhysRevLett.108.014301} {\bibfield  {journal} {\bibinfo
  {journal} {Physical Review Letters}\ }\textbf {\bibinfo {volume} {108}},\
  \bibinfo {pages} {1} (\bibinfo {year} {2012})}\BibitemShut {NoStop}%
\bibitem [{\citenamefont {Farhat}\ \emph {et~al.}(2010)\citenamefont {Farhat},
  \citenamefont {Guenneau}, \citenamefont {Enoch}, \citenamefont {Movchan},\
  and\ \citenamefont {Petursson}}]{Farhat2010}%
  \BibitemOpen
  \bibfield  {author} {\bibinfo {author} {\bibfnamefont {M.}~\bibnamefont
  {Farhat}}, \bibinfo {author} {\bibfnamefont {S.}~\bibnamefont {Guenneau}},
  \bibinfo {author} {\bibfnamefont {S.}~\bibnamefont {Enoch}}, \bibinfo
  {author} {\bibfnamefont {A.~B.}\ \bibnamefont {Movchan}}, \ and\ \bibinfo
  {author} {\bibfnamefont {G.~G.}\ \bibnamefont {Petursson}},\ }\href {\doibase
  10.1063/1.3327813} {\bibfield  {journal} {\bibinfo  {journal} {Applied
  Physics Letters}\ }\textbf {\bibinfo {volume} {96}},\ \bibinfo {pages}
  {081909} (\bibinfo {year} {2010})}\BibitemShut {NoStop}%
\bibitem [{\citenamefont {Pierre}\ \emph {et~al.}(2010)\citenamefont {Pierre},
  \citenamefont {Boyko}, \citenamefont {Belliard}, \citenamefont {Vasseur},\
  and\ \citenamefont {Bonello}}]{pierre2010negative}%
  \BibitemOpen
  \bibfield  {author} {\bibinfo {author} {\bibfnamefont {J.}~\bibnamefont
  {Pierre}}, \bibinfo {author} {\bibfnamefont {O.}~\bibnamefont {Boyko}},
  \bibinfo {author} {\bibfnamefont {L.}~\bibnamefont {Belliard}}, \bibinfo
  {author} {\bibfnamefont {J.}~\bibnamefont {Vasseur}}, \ and\ \bibinfo
  {author} {\bibfnamefont {B.}~\bibnamefont {Bonello}},\ }\href@noop {}
  {\bibfield  {journal} {\bibinfo  {journal} {Applied Physics Letters}\
  }\textbf {\bibinfo {volume} {97}},\ \bibinfo {pages} {121919} (\bibinfo
  {year} {2010})}\BibitemShut {NoStop}%
\bibitem [{\citenamefont {Wu}\ \emph {et~al.}(2011)\citenamefont {Wu},
  \citenamefont {Chen}, \citenamefont {Sun}, \citenamefont {Lin},\ and\
  \citenamefont {Huang}}]{lenteTTWu}%
  \BibitemOpen
  \bibfield  {author} {\bibinfo {author} {\bibfnamefont {T.-T.}\ \bibnamefont
  {Wu}}, \bibinfo {author} {\bibfnamefont {Y.-T.}\ \bibnamefont {Chen}},
  \bibinfo {author} {\bibfnamefont {J.-H.}\ \bibnamefont {Sun}}, \bibinfo
  {author} {\bibfnamefont {S.-C.~S.}\ \bibnamefont {Lin}}, \ and\ \bibinfo
  {author} {\bibfnamefont {T.~J.}\ \bibnamefont {Huang}},\ }\href {\doibase
  DOI:10.1063/1.3583660} {\bibfield  {journal} {\bibinfo  {journal} {App. Phys.
  Lett.}\ }\textbf {\bibinfo {volume} {98}},\ \bibinfo {pages} {171911}
  (\bibinfo {year} {2011})}\BibitemShut {NoStop}%
\bibitem [{\citenamefont {Zhao}\ \emph {et~al.}(2012)\citenamefont {Zhao},
  \citenamefont {Marchal}, \citenamefont {Bonello},\ and\ \citenamefont
  {Boyko}}]{zhao2012efficient}%
  \BibitemOpen
  \bibfield  {author} {\bibinfo {author} {\bibfnamefont {J.}~\bibnamefont
  {Zhao}}, \bibinfo {author} {\bibfnamefont {R.}~\bibnamefont {Marchal}},
  \bibinfo {author} {\bibfnamefont {B.}~\bibnamefont {Bonello}}, \ and\
  \bibinfo {author} {\bibfnamefont {O.}~\bibnamefont {Boyko}},\ }\href@noop {}
  {\bibfield  {journal} {\bibinfo  {journal} {Applied Physics Letters}\
  }\textbf {\bibinfo {volume} {101}},\ \bibinfo {pages} {261905} (\bibinfo
  {year} {2012})}\BibitemShut {NoStop}%
\bibitem [{\citenamefont {Climente}\ \emph {et~al.}(2014)\citenamefont
  {Climente}, \citenamefont {Torrent},\ and\ \citenamefont
  {S{\'a}nchez-Dehesa}}]{climente2014gradient}%
  \BibitemOpen
  \bibfield  {author} {\bibinfo {author} {\bibfnamefont {A.}~\bibnamefont
  {Climente}}, \bibinfo {author} {\bibfnamefont {D.}~\bibnamefont {Torrent}}, \
  and\ \bibinfo {author} {\bibfnamefont {J.}~\bibnamefont
  {S{\'a}nchez-Dehesa}},\ }\href@noop {} {\bibfield  {journal} {\bibinfo
  {journal} {Applied Physics Letters}\ }\textbf {\bibinfo {volume} {105}},\
  \bibinfo {pages} {064101} (\bibinfo {year} {2014})}\BibitemShut {NoStop}%
\bibitem [{\citenamefont {Krylov}(2012)}]{krylov2012resumen}%
  \BibitemOpen
  \bibfield  {author} {\bibinfo {author} {\bibfnamefont {V.}~\bibnamefont
  {Krylov}},\ }in\ \href@noop {} {\emph {\bibinfo {booktitle} {Proceedings of
  the International Conference on Noise and Vibration Engineering (ISMA
  2012)}}}\ (\bibinfo  {publisher} {Sas, P., Moens, D. and Jonckheer, S.
  (eds.).},\ \bibinfo {year} {2012})\ pp.\ \bibinfo {pages}
  {933--944}\BibitemShut {NoStop}%
\bibitem [{\citenamefont {Climente}\ \emph {et~al.}(2013)\citenamefont
  {Climente}, \citenamefont {Torrent},\ and\ \citenamefont
  {S{\'a}nchez-Dehesa}}]{climente2013omnidirectional}%
  \BibitemOpen
  \bibfield  {author} {\bibinfo {author} {\bibfnamefont {A.}~\bibnamefont
  {Climente}}, \bibinfo {author} {\bibfnamefont {D.}~\bibnamefont {Torrent}}, \
  and\ \bibinfo {author} {\bibfnamefont {J.}~\bibnamefont
  {S{\'a}nchez-Dehesa}},\ }\href@noop {} {\bibfield  {journal} {\bibinfo
  {journal} {Journal of Applied Physics}\ }\textbf {\bibinfo {volume} {114}},\
  \bibinfo {pages} {214903} (\bibinfo {year} {2013})}\BibitemShut {NoStop}%
\bibitem [{\citenamefont {Evans}\ and\ \citenamefont
  {Porter}(2007)}]{Evans2007}%
  \BibitemOpen
  \bibfield  {author} {\bibinfo {author} {\bibfnamefont {D.~V.}\ \bibnamefont
  {Evans}}\ and\ \bibinfo {author} {\bibfnamefont {R.}~\bibnamefont {Porter}},\
  }\href {\doibase 10.1007/s10665-006-9128-0} {\bibfield  {journal} {\bibinfo
  {journal} {Journal of Engineering Mathematics}\ }\textbf {\bibinfo {volume}
  {58}},\ \bibinfo {pages} {317} (\bibinfo {year} {2007})}\BibitemShut
  {NoStop}%
\bibitem [{\citenamefont {McPhedran}\ \emph {et~al.}(2009)\citenamefont
  {McPhedran}, \citenamefont {Movchan},\ and\ \citenamefont
  {Movchan}}]{mcphedran2009platonic}%
  \BibitemOpen
  \bibfield  {author} {\bibinfo {author} {\bibfnamefont {R.}~\bibnamefont
  {McPhedran}}, \bibinfo {author} {\bibfnamefont {A.}~\bibnamefont {Movchan}},
  \ and\ \bibinfo {author} {\bibfnamefont {N.}~\bibnamefont {Movchan}},\
  }\href@noop {} {\bibfield  {journal} {\bibinfo  {journal} {Mechanics of
  Materials}\ }\textbf {\bibinfo {volume} {41}},\ \bibinfo {pages} {356}
  (\bibinfo {year} {2009})}\BibitemShut {NoStop}%
\bibitem [{\citenamefont {Movchan}\ \emph {et~al.}(2007)\citenamefont
  {Movchan}, \citenamefont {Movchan},\ and\ \citenamefont
  {McPhedran}}]{Movchan2007}%
  \BibitemOpen
  \bibfield  {author} {\bibinfo {author} {\bibfnamefont {A.~B.}\ \bibnamefont
  {Movchan}}, \bibinfo {author} {\bibfnamefont {N.~V.}\ \bibnamefont
  {Movchan}}, \ and\ \bibinfo {author} {\bibfnamefont {R.~C.}\ \bibnamefont
  {McPhedran}},\ }\href {\doibase 10.1098/rspa.2007.1886} {\bibfield  {journal}
  {\bibinfo  {journal} {Proceedings of the Royal Society A: Mathematical,
  Physical and Engineering Sciences}\ }\textbf {\bibinfo {volume} {463}},\
  \bibinfo {pages} {2505} (\bibinfo {year} {2007})}\BibitemShut {NoStop}%
\bibitem [{\citenamefont {Pennec}\ \emph {et~al.}(2008)\citenamefont {Pennec},
  \citenamefont {Djafari-Rouhani}, \citenamefont {Larabi}, \citenamefont
  {Vasseur},\ and\ \citenamefont {Hladky-Hennion}}]{pennec2008low}%
  \BibitemOpen
  \bibfield  {author} {\bibinfo {author} {\bibfnamefont {Y.}~\bibnamefont
  {Pennec}}, \bibinfo {author} {\bibfnamefont {B.}~\bibnamefont
  {Djafari-Rouhani}}, \bibinfo {author} {\bibfnamefont {H.}~\bibnamefont
  {Larabi}}, \bibinfo {author} {\bibfnamefont {J.}~\bibnamefont {Vasseur}}, \
  and\ \bibinfo {author} {\bibfnamefont {A.}~\bibnamefont {Hladky-Hennion}},\
  }\href@noop {} {\bibfield  {journal} {\bibinfo  {journal} {Physical Review
  B}\ }\textbf {\bibinfo {volume} {78}},\ \bibinfo {pages} {104105} (\bibinfo
  {year} {2008})}\BibitemShut {NoStop}%
\bibitem [{\citenamefont {Pennec}\ \emph {et~al.}(2009)\citenamefont {Pennec},
  \citenamefont {Rouhani}, \citenamefont {Larabi}, \citenamefont {Akjouj},
  \citenamefont {Gillet}, \citenamefont {Vasseur},\ and\ \citenamefont
  {Thabet}}]{pennec2009phonon}%
  \BibitemOpen
  \bibfield  {author} {\bibinfo {author} {\bibfnamefont {Y.}~\bibnamefont
  {Pennec}}, \bibinfo {author} {\bibfnamefont {B.~D.}\ \bibnamefont {Rouhani}},
  \bibinfo {author} {\bibfnamefont {H.}~\bibnamefont {Larabi}}, \bibinfo
  {author} {\bibfnamefont {A.}~\bibnamefont {Akjouj}}, \bibinfo {author}
  {\bibfnamefont {J.}~\bibnamefont {Gillet}}, \bibinfo {author} {\bibfnamefont
  {J.}~\bibnamefont {Vasseur}}, \ and\ \bibinfo {author} {\bibfnamefont
  {G.}~\bibnamefont {Thabet}},\ }\href@noop {} {\bibfield  {journal} {\bibinfo
  {journal} {Physical Review B}\ }\textbf {\bibinfo {volume} {80}},\ \bibinfo
  {pages} {144302} (\bibinfo {year} {2009})}\BibitemShut {NoStop}%
\bibitem [{\citenamefont {Marchal}\ \emph {et~al.}(2012)\citenamefont
  {Marchal}, \citenamefont {Boyko}, \citenamefont {Bonello}, \citenamefont
  {Zhao}, \citenamefont {Belliard}, \citenamefont {Oudich}, \citenamefont
  {Pennec},\ and\ \citenamefont {Djafari-Rouhani}}]{marchal2012dynamics}%
  \BibitemOpen
  \bibfield  {author} {\bibinfo {author} {\bibfnamefont {R.}~\bibnamefont
  {Marchal}}, \bibinfo {author} {\bibfnamefont {O.}~\bibnamefont {Boyko}},
  \bibinfo {author} {\bibfnamefont {B.}~\bibnamefont {Bonello}}, \bibinfo
  {author} {\bibfnamefont {J.}~\bibnamefont {Zhao}}, \bibinfo {author}
  {\bibfnamefont {L.}~\bibnamefont {Belliard}}, \bibinfo {author}
  {\bibfnamefont {M.}~\bibnamefont {Oudich}}, \bibinfo {author} {\bibfnamefont
  {Y.}~\bibnamefont {Pennec}}, \ and\ \bibinfo {author} {\bibfnamefont
  {B.}~\bibnamefont {Djafari-Rouhani}},\ }\href@noop {} {\bibfield  {journal}
  {\bibinfo  {journal} {Physical Review B}\ }\textbf {\bibinfo {volume} {86}},\
  \bibinfo {pages} {224302} (\bibinfo {year} {2012})}\BibitemShut {NoStop}%
\bibitem [{\citenamefont {Xiao}\ \emph {et~al.}(2011)\citenamefont {Xiao},
  \citenamefont {Mace}, \citenamefont {Wen},\ and\ \citenamefont
  {Wen}}]{Xiao2011}%
  \BibitemOpen
  \bibfield  {author} {\bibinfo {author} {\bibfnamefont {Y.}~\bibnamefont
  {Xiao}}, \bibinfo {author} {\bibfnamefont {B.~R.}\ \bibnamefont {Mace}},
  \bibinfo {author} {\bibfnamefont {J.}~\bibnamefont {Wen}}, \ and\ \bibinfo
  {author} {\bibfnamefont {X.}~\bibnamefont {Wen}},\ }\href {\doibase
  10.1016/j.physleta.2011.02.044} {\bibfield  {journal} {\bibinfo  {journal}
  {Physics Letters A}\ }\textbf {\bibinfo {volume} {375}},\ \bibinfo {pages}
  {1485} (\bibinfo {year} {2011})}\BibitemShut {NoStop}%
\bibitem [{\citenamefont {Xiao}\ \emph {et~al.}(2012)\citenamefont {Xiao},
  \citenamefont {Wen},\ and\ \citenamefont {Wen}}]{Xiao2012}%
  \BibitemOpen
  \bibfield  {author} {\bibinfo {author} {\bibfnamefont {Y.}~\bibnamefont
  {Xiao}}, \bibinfo {author} {\bibfnamefont {J.}~\bibnamefont {Wen}}, \ and\
  \bibinfo {author} {\bibfnamefont {X.}~\bibnamefont {Wen}},\ }\href {\doibase
  10.1088/0022-3727/45/19/195401} {\bibfield  {journal} {\bibinfo  {journal}
  {Journal of Physics D: Applied Physics}\ }\textbf {\bibinfo {volume} {45}},\
  \bibinfo {pages} {195401} (\bibinfo {year} {2012})}\BibitemShut {NoStop}%
\bibitem [{\citenamefont {Torrent}\ \emph {et~al.}(2013)\citenamefont
  {Torrent}, \citenamefont {Mayou},\ and\ \citenamefont
  {Sanchez-Dehesa}}]{Torrent2013Graphene}%
  \BibitemOpen
  \bibfield  {author} {\bibinfo {author} {\bibfnamefont {D.}~\bibnamefont
  {Torrent}}, \bibinfo {author} {\bibfnamefont {D.}~\bibnamefont {Mayou}}, \
  and\ \bibinfo {author} {\bibfnamefont {J.}~\bibnamefont {Sanchez-Dehesa}},\
  }\href {\doibase {10.1103/PhysRevB.87.115143}} {\bibfield  {journal}
  {\bibinfo  {journal} {{PHYSICAL REVIEW B}}\ }\textbf {\bibinfo {volume}
  {{87}}} (\bibinfo {year} {{2013}}),\
  {10.1103/PhysRevB.87.115143}}\BibitemShut {NoStop}%
\bibitem [{\citenamefont {Zhu}\ \emph {et~al.}(2012)\citenamefont {Zhu},
  \citenamefont {Liu}, \citenamefont {Huang}, \citenamefont {Huang},\ and\
  \citenamefont {Sun}}]{Zhu2012}%
  \BibitemOpen
  \bibfield  {author} {\bibinfo {author} {\bibfnamefont {R.}~\bibnamefont
  {Zhu}}, \bibinfo {author} {\bibfnamefont {X.~N.}\ \bibnamefont {Liu}},
  \bibinfo {author} {\bibfnamefont {G.~L.}\ \bibnamefont {Huang}}, \bibinfo
  {author} {\bibfnamefont {H.~H.}\ \bibnamefont {Huang}}, \ and\ \bibinfo
  {author} {\bibfnamefont {C.~T.}\ \bibnamefont {Sun}},\ }\href {\doibase
  10.1103/PhysRevB.86.144307} {\bibfield  {journal} {\bibinfo  {journal}
  {Physical Review B}\ }\textbf {\bibinfo {volume} {86}},\ \bibinfo {pages}
  {144307} (\bibinfo {year} {2012})}\BibitemShut {NoStop}%
\bibitem [{\citenamefont {Vynck}\ \emph {et~al.}(2009)\citenamefont {Vynck},
  \citenamefont {Felbacq}, \citenamefont {Centeno}, \citenamefont
  {C{\u{a}}buz}, \citenamefont {Cassagne},\ and\ \citenamefont
  {Guizal}}]{vynck2009all}%
  \BibitemOpen
  \bibfield  {author} {\bibinfo {author} {\bibfnamefont {K.}~\bibnamefont
  {Vynck}}, \bibinfo {author} {\bibfnamefont {D.}~\bibnamefont {Felbacq}},
  \bibinfo {author} {\bibfnamefont {E.}~\bibnamefont {Centeno}}, \bibinfo
  {author} {\bibfnamefont {A.}~\bibnamefont {C{\u{a}}buz}}, \bibinfo {author}
  {\bibfnamefont {D.}~\bibnamefont {Cassagne}}, \ and\ \bibinfo {author}
  {\bibfnamefont {B.}~\bibnamefont {Guizal}},\ }\href@noop {} {\bibfield
  {journal} {\bibinfo  {journal} {Physical review letters}\ }\textbf {\bibinfo
  {volume} {102}},\ \bibinfo {pages} {133901} (\bibinfo {year}
  {2009})}\BibitemShut {NoStop}%
\bibitem [{\citenamefont {Torrent}\ and\ \citenamefont
  {S{\'a}nchez-Dehesa}(2011)}]{torrent2011multiple}%
  \BibitemOpen
  \bibfield  {author} {\bibinfo {author} {\bibfnamefont {D.}~\bibnamefont
  {Torrent}}\ and\ \bibinfo {author} {\bibfnamefont {J.}~\bibnamefont
  {S{\'a}nchez-Dehesa}},\ }\href@noop {} {\bibfield  {journal} {\bibinfo
  {journal} {New Journal of Physics}\ }\textbf {\bibinfo {volume} {13}},\
  \bibinfo {pages} {093018} (\bibinfo {year} {2011})}\BibitemShut {NoStop}%
\bibitem [{\citenamefont {Wu}\ \emph {et~al.}(2007)\citenamefont {Wu},
  \citenamefont {Lai},\ and\ \citenamefont {Zhang}}]{wu2007effective}%
  \BibitemOpen
  \bibfield  {author} {\bibinfo {author} {\bibfnamefont {Y.}~\bibnamefont
  {Wu}}, \bibinfo {author} {\bibfnamefont {Y.}~\bibnamefont {Lai}}, \ and\
  \bibinfo {author} {\bibfnamefont {Z.-Q.}\ \bibnamefont {Zhang}},\ }\href@noop
  {} {\bibfield  {journal} {\bibinfo  {journal} {Physical Review B}\ }\textbf
  {\bibinfo {volume} {76}},\ \bibinfo {pages} {205313} (\bibinfo {year}
  {2007})}\BibitemShut {NoStop}%
\bibitem [{\citenamefont {Zhou}\ and\ \citenamefont
  {Hu}(2009)}]{zhou2009analytic}%
  \BibitemOpen
  \bibfield  {author} {\bibinfo {author} {\bibfnamefont {X.}~\bibnamefont
  {Zhou}}\ and\ \bibinfo {author} {\bibfnamefont {G.}~\bibnamefont {Hu}},\
  }\href@noop {} {\bibfield  {journal} {\bibinfo  {journal} {Physical Review
  B}\ }\textbf {\bibinfo {volume} {79}},\ \bibinfo {pages} {195109} (\bibinfo
  {year} {2009})}\BibitemShut {NoStop}%
\bibitem [{\citenamefont {Graff}(1991)}]{Graff}%
  \BibitemOpen
  \bibfield  {author} {\bibinfo {author} {\bibfnamefont {K.~F.}\ \bibnamefont
  {Graff}},\ }\href@noop {} {\emph {\bibinfo {title} {Wave Motion in elastic
  solids, 2nd Ed}}}\ (\bibinfo  {publisher} {Dover},\ \bibinfo {year}
  {1991})\BibitemShut {NoStop}%
\bibitem [{\citenamefont {Timoshenko}(1940)}]{Timoshenko}%
  \BibitemOpen
  \bibfield  {author} {\bibinfo {author} {\bibfnamefont {S.}~\bibnamefont
  {Timoshenko}},\ }\href@noop {} {\emph {\bibinfo {title} {Theory of Plates and
  Shells}}}\ (\bibinfo  {publisher} {Mc. Graw-Hill},\ \bibinfo {year}
  {1940})\BibitemShut {NoStop}%
\bibitem [{\citenamefont {Torrent}\ and\ \citenamefont
  {S\'anchez-Dehesa}(2006)}]{homoDani2}%
  \BibitemOpen
  \bibfield  {author} {\bibinfo {author} {\bibfnamefont {D.}~\bibnamefont
  {Torrent}}\ and\ \bibinfo {author} {\bibfnamefont {J.}~\bibnamefont
  {S\'anchez-Dehesa}},\ }\href@noop {} {\bibfield  {journal} {\bibinfo
  {journal} {Phys. Rev. B}\ }\textbf {\bibinfo {volume} {74}},\ \bibinfo
  {pages} {224305} (\bibinfo {year} {2006})}\BibitemShut {NoStop}%
\bibitem [{\citenamefont {Waterman}(1965)}]{waterman1965matrix}%
  \BibitemOpen
  \bibfield  {author} {\bibinfo {author} {\bibfnamefont {P.}~\bibnamefont
  {Waterman}},\ }\href@noop {} {\bibfield  {journal} {\bibinfo  {journal}
  {Proceedings of the IEEE}\ }\textbf {\bibinfo {volume} {53}},\ \bibinfo
  {pages} {805} (\bibinfo {year} {1965})}\BibitemShut {NoStop}%
\bibitem [{\citenamefont {Waterman}(2005)}]{waterman2005new}%
  \BibitemOpen
  \bibfield  {author} {\bibinfo {author} {\bibfnamefont {P.}~\bibnamefont
  {Waterman}},\ }\href@noop {} {\bibfield  {journal} {\bibinfo  {journal} {The
  journal of the acoustical society of America}\ }\textbf {\bibinfo {volume}
  {45}},\ \bibinfo {pages} {1417} (\bibinfo {year} {2005})}\BibitemShut
  {NoStop}%
\bibitem [{\citenamefont {Felbacq}\ \emph {et~al.}(1994)\citenamefont
  {Felbacq}, \citenamefont {Tayeb},\ and\ \citenamefont
  {Maystre}}]{felbacq1994scattering}%
  \BibitemOpen
  \bibfield  {author} {\bibinfo {author} {\bibfnamefont {D.}~\bibnamefont
  {Felbacq}}, \bibinfo {author} {\bibfnamefont {G.}~\bibnamefont {Tayeb}}, \
  and\ \bibinfo {author} {\bibfnamefont {D.}~\bibnamefont {Maystre}},\
  }\href@noop {} {\bibfield  {journal} {\bibinfo  {journal} {JOSA A}\ }\textbf
  {\bibinfo {volume} {11}},\ \bibinfo {pages} {2526} (\bibinfo {year}
  {1994})}\BibitemShut {NoStop}%
\bibitem [{\citenamefont {Torrent}\ \emph {et~al.}(2006)\citenamefont
  {Torrent}, \citenamefont {H\aa{}kansson}, \citenamefont {Cervera},\ and\
  \citenamefont {S\'anchez-Dehesa}}]{homoDani1}%
  \BibitemOpen
  \bibfield  {author} {\bibinfo {author} {\bibfnamefont {D.}~\bibnamefont
  {Torrent}}, \bibinfo {author} {\bibfnamefont {A.}~\bibnamefont
  {H\aa{}kansson}}, \bibinfo {author} {\bibfnamefont {F.}~\bibnamefont
  {Cervera}}, \ and\ \bibinfo {author} {\bibfnamefont {J.}~\bibnamefont
  {S\'anchez-Dehesa}},\ }\href@noop {} {\bibfield  {journal} {\bibinfo
  {journal} {Phys. Rev. Lett.}\ }\textbf {\bibinfo {volume} {96}},\ \bibinfo
  {pages} {204302} (\bibinfo {year} {2006})}\BibitemShut {NoStop}%
\bibitem [{\citenamefont {Torrent}\ \emph {et~al.}(2007)\citenamefont
  {Torrent}, \citenamefont {S{\'a}nchez-Dehesa},\ and\ \citenamefont
  {Cervera}}]{torrent2007evidence}%
  \BibitemOpen
  \bibfield  {author} {\bibinfo {author} {\bibfnamefont {D.}~\bibnamefont
  {Torrent}}, \bibinfo {author} {\bibfnamefont {J.}~\bibnamefont
  {S{\'a}nchez-Dehesa}}, \ and\ \bibinfo {author} {\bibfnamefont
  {F.}~\bibnamefont {Cervera}},\ }\href@noop {} {\bibfield  {journal} {\bibinfo
   {journal} {Physical Review B}\ }\textbf {\bibinfo {volume} {75}},\ \bibinfo
  {pages} {241404} (\bibinfo {year} {2007})}\BibitemShut {NoStop}%
\bibitem [{\citenamefont {Norris}\ and\ \citenamefont
  {Vemula}(1995)}]{NorrisVemula}%
  \BibitemOpen
  \bibfield  {author} {\bibinfo {author} {\bibfnamefont {A.}~\bibnamefont
  {Norris}}\ and\ \bibinfo {author} {\bibfnamefont {C.}~\bibnamefont
  {Vemula}},\ }\href@noop {} {\bibfield  {journal} {\bibinfo  {journal}
  {Journal of Sound and Vibration}\ }\textbf {\bibinfo {volume} {181}},\
  \bibinfo {pages} {115 } (\bibinfo {year} {1995})}\BibitemShut {NoStop}%
\bibitem [{\citenamefont {Parnell}\ and\ \citenamefont
  {Martin}(2011)}]{Parnell2011}%
  \BibitemOpen
  \bibfield  {author} {\bibinfo {author} {\bibfnamefont {W.}~\bibnamefont
  {Parnell}}\ and\ \bibinfo {author} {\bibfnamefont {P.}~\bibnamefont
  {Martin}},\ }\href {\doibase 10.1016/j.wavemoti.2010.10.004} {\bibfield
  {journal} {\bibinfo  {journal} {Wave Motion}\ }\textbf {\bibinfo {volume}
  {48}},\ \bibinfo {pages} {161} (\bibinfo {year} {2011})}\BibitemShut
  {NoStop}%
\bibitem [{\citenamefont {Lee}\ and\ \citenamefont
  {Chen}(2010)}]{lee2010scattering}%
  \BibitemOpen
  \bibfield  {author} {\bibinfo {author} {\bibfnamefont {W.-M.}\ \bibnamefont
  {Lee}}\ and\ \bibinfo {author} {\bibfnamefont {J.-T.}\ \bibnamefont {Chen}},\
  }\href@noop {} {\bibfield  {journal} {\bibinfo  {journal} {International
  Journal of Solids and Structures}\ }\textbf {\bibinfo {volume} {47}},\
  \bibinfo {pages} {1118} (\bibinfo {year} {2010})}\BibitemShut {NoStop}%
\bibitem [{\citenamefont {Martin}\ \emph {et~al.}(2010)\citenamefont {Martin},
  \citenamefont {Nicholas}, \citenamefont {Orris}, \citenamefont {Cai},
  \citenamefont {Torrent},\ and\ \citenamefont {Sanchez-Dehesa}}]{Martin2010}%
  \BibitemOpen
  \bibfield  {author} {\bibinfo {author} {\bibfnamefont {T.}~\bibnamefont
  {Martin}}, \bibinfo {author} {\bibfnamefont {M.}~\bibnamefont {Nicholas}},
  \bibinfo {author} {\bibfnamefont {G.}~\bibnamefont {Orris}}, \bibinfo
  {author} {\bibfnamefont {L.-W.}\ \bibnamefont {Cai}}, \bibinfo {author}
  {\bibfnamefont {D.}~\bibnamefont {Torrent}}, \ and\ \bibinfo {author}
  {\bibfnamefont {J.}~\bibnamefont {Sanchez-Dehesa}},\ }\href
  {http://www.scopus.com/inward/record.url?eid=2-s2.0-77956841493&partnerID=40&md5=195c80293a95411898ee6b6e6cdfd6e1}
  {\bibfield  {journal} {\bibinfo  {journal} {Applied Physics Letters}\
  }\textbf {\bibinfo {volume} {97}} (\bibinfo {year} {2010})}\BibitemShut
  {NoStop}%
\bibitem [{\citenamefont {Climente}\ \emph {et~al.}(2010)\citenamefont
  {Climente}, \citenamefont {Torrent},\ and\ \citenamefont
  {Sanchez-Dehesa}}]{Climente2010}%
  \BibitemOpen
  \bibfield  {author} {\bibinfo {author} {\bibfnamefont {A.}~\bibnamefont
  {Climente}}, \bibinfo {author} {\bibfnamefont {D.}~\bibnamefont {Torrent}}, \
  and\ \bibinfo {author} {\bibfnamefont {J.}~\bibnamefont {Sanchez-Dehesa}},\
  }\href
  {http://www.scopus.com/inward/record.url?eid=2-s2.0-77956581113&partnerID=40&md5=917f078a1d3bd0a02be76f33b267c05f}
  {\bibfield  {journal} {\bibinfo  {journal} {Applied Physics Letters}\
  }\textbf {\bibinfo {volume} {97}} (\bibinfo {year} {2010})}\BibitemShut
  {NoStop}%
\bibitem [{\citenamefont {Climente}\ \emph {et~al.}(2012)\citenamefont
  {Climente}, \citenamefont {Torrent},\ and\ \citenamefont
  {Sanchez-Dehesa}}]{Climente2012}%
  \BibitemOpen
  \bibfield  {author} {\bibinfo {author} {\bibfnamefont {A.}~\bibnamefont
  {Climente}}, \bibinfo {author} {\bibfnamefont {D.}~\bibnamefont {Torrent}}, \
  and\ \bibinfo {author} {\bibfnamefont {J.}~\bibnamefont {Sanchez-Dehesa}},\
  }\href
  {http://www.scopus.com/inward/record.url?eid=2-s2.0-84859805870&partnerID=40&md5=2561bf2e211d79b92179fdd29df660c4}
  {\bibfield  {journal} {\bibinfo  {journal} {Applied Physics Letters}\
  }\textbf {\bibinfo {volume} {100}} (\bibinfo {year} {2012})}\BibitemShut
  {NoStop}%
\end{thebibliography}
%

\end{document}